\def\bm#1{\mbox{\boldmath $#1$}}
\documentstyle[12pt]{article}

\begin{document}

\hoffset = -1truecm
\voffset = -3truecm

\thispagestyle{empty}
\begin{flushright}
{ \bf DFTUZ/95/06}\\
{\bf hep-th/9505117}
\end{flushright}
$\ $
\vskip 3truecm

\begin{center}

{ \large \bf  ANYONS AS SPINNING PARTICLES}\\
\vskip1.0cm
{ \bf
Jos\'e L. Cort\'es\footnote{E-mail: cortes@cc.unizar.es}
 and Mikhail S. Plyushchay\footnote{On leave from the
{\it Institute for High Energy Physics,
Protvino, Moscow Region, Russia};
e-mail: mikhail@cc.unizar.es}}\\[0.3cm]
{\it Departamento de F\'{\i}sica Te\'orica,
Facultad de Ciencias}\\
{\it Universidad de Zaragoza, 50009 Zaragoza, Spain}\\[0.5cm]
\end{center}

\vskip2.0cm
\begin{center}
                            {\bf Abstract}
\end{center}
A model-independent formulation of anyons as spinning particles
is presented.  The general properties of the classical theory of
(2+1)-dimensional relativistic fractional spin particles and
some properties of their quantum theory are investigated. The
relationship between all the known approaches to anyons as
spinning particles is established.  Some widespread misleading
notions on the general properties of (2+1)-dimensional anyons
are removed.
\vskip1.0cm
\begin{center}
To be published in {\bf Int. J. Mod. Phys. A}
\end{center}

\makeatletter
\@addtoreset{equation}{section}
  \def\theequation{\thesection.\arabic{equation}}
\makeatother

\newpage
\section{Introduction}

It is now well known that the case of (2+1)-dimensional space-time is a
special one from the point of view of spinning particles: only in this case
spin can take arbitrary values on the real line \cite{schon,binegar}.  Such
particles with fractional (arbitrary) spin, called anyons, satisfy an
exotic fractional statistics, realizing representations of the braid group
\cite{leimyr}--\cite{wil}. It is believed that anyons in the form of
quasiparticles are related to the planar physical phenomena such as
fractional quantum Hall effect and high-$T_c$ superconductivity
\cite{appl}.

The theory of anyons as spinning particles has specific features just
from the point of view of classical mechanics. It is obvious that
fractional spin particles cannot be described at the classical level by the
standard pseudoclassical approach which uses Grassmann variables for taking
into account spin degrees of freedom \cite{pseudo}.  This is due to the
fact that Grassmann variables lead after quantization to the
finite-dimensional representations of the (2+1)-dimensional Lorentz group.
Such representations correspond only to the cases of integer or
half-integer spin \cite{cor}, and, therefore, the description of anyons
requires some nonstandard approaches.
Nevertheless,
the first model of relativistic particle with fractional
(arbitrary) spin was constructed in ref. \cite{pl1} on the basis
of an analogy with the pseudoclassical approach \cite{pl2}.
In that model spin degrees of freedom were described by
commuting translationally invariant variables and configuration
space of the system had a nontrivial topology corresponding to
the topology of the simplest nonrelativistic anyon system
\cite{wilmac}.

Another attempt to construct a model of
relativistic fractional spin particle was undertaken in ref.
\cite{gerb} following the (3+1)-dimensional approach of
Balachandran et al.  \cite{bal}.  The paper \cite{gerb}
comprised the discussion of the classical theory and contained some
general statements on the properties of (2+1)-dimensional
relativistic spinning particles.  First such a statement is that
in the presence of spin the angular momentum/boost vector of a
particle has a contribution proportional to its momentum.  This
contribution must, according to the paper \cite{gerb}, satisfy
the algebra of the (2+1)-dimensional Lorentz group ${\rm SO(2,1)}$
at the level of Poisson brackets.  These two statements
being correct only in some special cases but misleading in
the general case, became, unfortunately, popular notions
and were reproduced noncritically in a series of papers on the
subject.  In particular, the parallelness of the spin
contribution to the energy-momentum vector was declared in ref.
\cite{chou1}, devoted to the electromagnetic interaction of
anyons, as to be a fundamental property for spinning
particles in 2+1 dimensions.  Moreover, the attempt to
incorporate both general statements of ref.  \cite{gerb} led
to the construction of the self-contradicting model \cite{chai}.
Subsequently, that model was criticized in the literature (see,
e.g., ref. \cite{jn2}), but the above-mentioned basic misleading
statements were not removed.  In turn, the paper \cite{chai}
itself contained another misleading statement, that the second
time derivative of the coordinates of a free relativistic
spinning particle has to be equal to zero.  This last statement
was reproduced in the recent paper \cite{dutra}, where it was
also declared that it is not possible to keep simultaneously the
free nature of the anyon and the translational invariance.

The purpose of the present paper is to investigate the general properties of
the classical theory of relativistic fractional spin particles and some
properties of the corresponding quantum theory.  This, in particular, will
allow us to demonstrate that the above-mentioned widespread
notions on (2+1)-dimensional anyons
are incorrect in the general case, and, therefore, can, as it has
been stated above, be misleading.  The analysis is based on
the Poincar\'e group being the exact symmetry group of the
theory, and, so, it will also give us a possibility to
demonstrate that the general statement of ref. \cite{dutra} is
incorrect.

Our consideration includes all the known approaches to the description of
anyons as (2+1)-dimensional spinning particles.  Such approaches do not use
Chern-Simons U(1) gauge field constructions taking place in another, more
developed approach to anyons \cite{sem,sred,appl},
and they can be classified in
the following way.  The essential feature of the first above-mentioned
model \cite{pl1} is that in the case of quantizing the system by the
reduced phase space quantization method, spin is present in it as an
independent operator.  As a result, fractional spin states are described
there by means of multi-valued wave functions in the case of the standard
realization of the spin operator or by single-valued functions in the case
of its nonstandard realization (see below, section 6).  Two analogous
possibilities with single- and multi-valued wave functions take also place
in the case of the covariant quantization of the model by the Dirac method
\cite{pl1}.  Another model of such type was proposed in ref.
\cite{pl3} and subsequently analysed in ref. \cite{pl4}, where
some very superficial speculations were given
on the relation of the approach of the models \cite{pl1,pl3}
to another possible approach to anyons which uses
infinite-dimensional representations of the universal covering group ${\rm
\overline{SO(2,1)}}$ of the (2+1)-dimensional Lorentz group SO(2,1) (or
${\rm \overline{SL(2,R)}}$, equivalent to it)
\cite{cor},\cite{pl3}--\cite{pl6}.  This latter
approach ascends
in its origin to the (2+1)-dimensional model of the relativistic
particle with torsion \cite{pl5,polyakov}.
At last,
fractional spin particles can be described proceeding from the
nontrivial symplectic two-form corresponding to the
monopole-like Poisson structure on the phase space of
a relativistic particle \cite{jn1}.  For the first time such a
possibility to describe anyons was, probably, pointed out (indirectly) in
ref. \cite{schon}
(see also ref. \cite{ss} and relevant papers \cite{abgss}).
Recently, it was exploited in refs. \cite{chou1,chou2} for
introducing the interaction of anyons with electromagnetic and
gravitational fields.

The present investigation will result in establishing
the relationship between all the listed approaches to anyons as
spinning particles.

The paper is organized as follows.  In section 2 we ``derive"
two different but related possibilities for describing
relativistic massive particles with an arbitrary fixed spin
$s\neq 0$ within the framework of the classical canonical
approach.  It will be done proceeding from the general
properties of the classical Poincar\'e group and exploiting an
observation of a similarity, peculiar to the (2+1)-dimensional
space-time, of the general case, characterized by $s\neq 0$, to
the spinless case, $s=0$.  Here we take into account the Jacobi
identities for independent phase space variables of the system,
that allows us to generalize our preliminary analysis of ref.
\cite{corpl} to the most general case.  One of the two
possibilities, called by us as a minimal formulation (approach),
consists in using for the purpose only the coordinate and
momentum phase space variables and in fixing the value of
nontrivial spin in a strong sense.  Another possibility, called
an extended formulation, uses auxiliary internal phase space
variables and fixes spin in a weak sense, with the help of the
corresponding spin constraint.  The approach of refs.
\cite{chou1,jn1,ss,chou2} with nontrivial symplectic
two-form corresponds to the minimal formulation and is contained
in it as a special, covariant, particular case. The extended
formulation comprises two other above-mentioned approaches to
anyons as spinning particles.

Section 3 is devoted to the investigation of the minimal approach.  Here we
shall find the general form of the brackets for the coordinates of the
particle and the corresponding form for the total angular momentum vector,
and, moreover, investigate the Lorentz properties of the coordinates. Then we
shall consider two special cases in more detail. The first one comprises
covariant coordinates being nonlocalizable in the sense of brackets,
whereas the second case is given by localizable coordinates, having trivial
brackets between different components, but being noncovariant ones from the
point of view of their Lorentz properties.  We shall establish the connection
between these two cases and use it for realizing the quantization procedure
for the minimal canonical approach.

Section 4 is devoted to the consideration of the general properties of the
extended canonical formulation and its relationship to the minimal one. The
concrete models corresponding to the extended formulation are considered in
two subsequent sections.  In section 5 we analyse in detail the model
proposed by us in paper \cite{corpl}. We shall discuss this model within
the framework of the lagrangian and hamiltonian approaches, and construct
the general solutions to its equations of motion. This, in particular, will
allow us to demonstrate that the classical analog
of the quantum Zitterbewegung takes place in the general case here,
and, as a consequence, the above-mentioned statement of refs.
\cite{chai,dutra} on the second derivative of the coordinates
of a free relativistic spinning particle in 2+1 dimensions
turns out generally to be incorrect.  We shall conclude this
section by discussing the quantization of the model.  At the
classical level, the model contains two cases with essentially
different topology of the internal phase subspace of the system
being a two-sheet or one-sheet hyperboloid.  Correspondingly,
the quantum theory comprises as an essential ingredient the
infinite-dimensional representations of the ${\rm
\overline{SL(2,R)}}$ group either of the so called discrete type series
(half-bounded representations) or of the continuous series (unbounded
representations).

In section 6, the model of refs. \cite{pl3,pl4} is considered.
We shall get another form for its lagrangian and with the help of it analyse
the classical theory of the model. Then we shall discuss the quantum theory
revealing its analogy with that for the model from section 5.  This
observation together with the general results obtained in section 4 will help
us to establish the one-to-one correspondence of the reduced phase
space description of the model with the
canonical description of the model from section 5
for the case corresponding
to the continuous series of representations of ${\rm
\overline{SL(2,R)}}$ at the quantum level. Thus, we shall find that these two
extended models together with corresponding quantum schemes
turn out to be closely related. In conclusion of this section
we shall discuss a possible interpretation of the model ascending to the
original analysis of Leinaas and Myrheim on the topology of configuration
space of the systems of identical particles, which resulted in
their discovery of fractional statistics \cite{leimyr}.
Moreover, here we shall comment upon the relationship of the
model to the approach of Balachandran et al. \cite{bal}.

Section 7 comprises a summary of the paper and a list of the
main results removing misleading notions on the general
properties of anyons.

\section{Spin in 2+1 dimensions}

In classical mechanics, the algebra of (2+1)-dimensional
Poincar\'e group ISO(2,1)
is given by the Poisson brackets of its generators being the energy-momentum
vector $p_{\mu}$ and the total angular momentum vector
${\cal J}_{\mu}$ dual to the
total angular momentum tensor ${\cal J}^{\nu\lambda}$,
${\cal J}_{\mu}=-\frac{1}{2}\epsilon_{\mu\nu\lambda}{\cal J}^{\nu\lambda}$,
\begin{eqnarray}
\{p_{\mu},p_{\nu}\}&=&0,
\label{pp}\\
\{{\cal J}_{\mu},{\cal J}_{\nu}\}
&=&-\epsilon_{\mu\nu\lambda}{\cal J}^{\lambda},
\label{caljj}\\
\{{\cal J}_{\mu},p_{\nu}\}&=&-\epsilon_{\mu\nu\lambda}p^{\lambda},
\label{caljp}
\end{eqnarray}
where we have used the metric $\eta_{\mu\nu}={\rm diag} (-,+,+)$ and a
totally antisymmetric tensor $\epsilon_{\mu\nu\lambda}$, $\epsilon^{012}=1$.
The components of the total angular momentum vector ${\cal J}^{\mu}$ are the
generators of the space rotations ($\mu=0$) and Lorentz boosts ($\mu=1,2$),
whereas the energy-momentum vector $p^{\mu}$ generates the
space-time translations.

The quantities $p^{\mu}p_{\mu}$ and $p^{\mu}{\cal J}_{\mu}$ lie in the center
of algebra (\ref{pp})--(\ref{caljp}). So, in the quantum case their analogs
are the Casimir operators of the quantum mechanical Poincar\'e group ${\rm
\overline{ISO(2,1)}}$ with eigenvalues characterizing its irreducible
representations. In the massive case, $-p^{2}=m^{2}>0$, the pseudoscalar
quantity
\begin{equation}
S=\frac{p^{\mu}{\cal J}_{\mu}}{\sqrt{-p^{2}}}
\label{spin}
\end{equation}
has a sense of spin and the eigenvalues of its quantum analog can take any
values $s\in {\bm R}$ \cite{binegar}.  Moreover, we have here the following
specific situation due to the (pseudo)-scalar nature of spin.  In
(2+1)-dimensional space-time a relativistic particle with fixed mass $m>0$
and fixed spin $s\neq 0$ has the same number of degrees of freedom as the
relativistic massive scalar particle with zero spin.  Proceeding from this
observation, we pass over to the investigation of possible approaches for
describing relativistic massive particle of an arbitrary (fixed) spin.

Within the framework of the  canonical approach such a system can be
described in the following way.
Introduce coordinates of the particle $x_{\mu}$, $\mu=0,1,2$,
conjugate to the momenta $p^{\mu}$,
\begin{equation}
\{x_{\mu},p_{\nu}\}=\eta_{\mu\nu}.
\label{xp}
\end{equation}
These brackets mean that momenta $p^{\mu}$ simultaneously are the
generators of the space-time translations.
The brackets for the coordinates themselves we denote as
\begin{equation}
\{x_{\mu},x_{\nu}\}=\epsilon_{\mu\nu\lambda}R^{\lambda},
\label{xxr}
\end{equation}
having in mind that $R^{\mu}$ can be some function of $x_{\mu}$ and
$p_{\mu}$, and, possibly, of other (auxiliary) variables.  Its admissible
structure will be analysed below.

Now, let us introduce the constraint
\begin{equation}
p^{2}+m^{2}\approx 0.
\label{p2m2}
\end{equation}
This constraint fixes the mass in a weak sense \cite{sunder}, and its quantum
analog is nothing else as the Klein-Gordon equation.  The total angular
momentum for a spinning particle can be taken in the form generalizing that
for the spinless (scalar) particle:
\begin{equation}
{\cal J}_{\mu}=-\epsilon_{\mu\nu\lambda}x^{\nu}p^{\lambda} +J_{\mu}.
\label{calj}
\end{equation}
The second term $J_{\mu}$ in eq. (\ref{calj}) takes into
account spin $s\neq 0$.  This can be done in two
different ways. One possibility consists in introducing some auxiliary
internal phase space
variables $z_n$, $n=1,\ldots,2N$, independent of external variables
$x_{\mu}$ and $p_{\mu}$, $\{z_{n},x_{\mu}\}=\{z_{n},p_{\mu}\}=0$, and in
supplementing the mass shell constraint (\ref{p2m2}) with the constraint
\begin{equation}
pJ -sm\approx 0.
\label{pjsm}
\end{equation}
Here we suppose that $J_{\mu}$ depends on these internal variables.
Relations  (\ref{p2m2}), (\ref{calj})
and definition (\ref{spin}) mean that
this constraint fixes the value of the particle spin.
Due to independence of $J_\mu$ on $x_\mu$ (see below),
the constraints (\ref{pjsm}) and
(\ref{p2m2}) form
the trivial algebra of the first class constraints.
Moreover, in order to have a pure spin system, without any other internal
degrees of freedom, we must supplement constraint (\ref{pjsm}) with
a corresponding number of first and/or second class constraints to eliminate
the remaining $N-1$ internal phase space degrees of freedom ($2(N-1)$
variables). As a result, the system will have the same number of degrees of
freedom as the relativistic spinless particle in 2+1 dimensions.

Another possibility consists in fixing the value of spin in a strong sense,
without introducing any auxiliary internal degrees of freedom and
corresponding constraints freezing these degrees of freedom. It can be done
by choosing the second term in the total angular momentum (\ref{calj}) in the
form
\begin{equation}
J_{\mu}=-se^{(0)}_{\mu}+J^{\bot}_{\mu},
\label{mini}
\end{equation}
where $e^{(0)}_{\mu}=p_{\mu}/\sqrt{-p^{2}}$ and $J^{\bot}_{\mu}$ is
transverse to $p^{\mu}$, $J^{\bot}_{\mu}p^{\mu}=0$.  Indeed, in this case the
system will have the same number of degrees of freedom as a scalar massive
particle has.  Moreover, if we shall satisfy relations (\ref{caljj}) and
(\ref{caljp}), then representation (\ref{mini}) together with eqs.
(\ref{spin}) and (\ref{p2m2}) will guarantee that we have a relativistic
massive particle with nontrivial spin $s$.  We shall call
this latter possibility for the
description of fractional (arbitrary) spin particles
the minimal approach (formulation),
whereas the former one will be called the extended approach.

The vector $e^{(0)}_{\mu}$, introduced above, can be supplemented with
the pair of momentum-dependent objects
$e^{(i)}_{\mu}=e^{(i)}_{\mu}(p)$, $i=1,2,$ so that the three
$e^{(\alpha)}_{\mu}$, $\alpha=0,1,2,$ will form a complete triad:
\begin{equation}
e^{(\alpha)}_{\mu}\eta_{\alpha\beta}e^{(\beta)}_{\nu}=\eta_{\mu\nu},\quad
e^{(\alpha)}_{\mu}\eta^{\mu\nu}e^{(\beta)}_{\nu}=\eta^{\alpha\beta},
\label{triad}
\end{equation}
whose orientation can be chosen in different ways, in particular, we
can fix it as
\[
\epsilon_{\mu\nu\lambda}e^{(0)\mu}e^{(1)\nu}e^{(2)\lambda}=1.
\]
Such a triad will be used in a further analysis.

Before passing over to the analysis of the two described possibilities
and their relationship, let us make some general observations,
which will be necessary for a subsequent consideration.
First of all, we note that the brackets (\ref{caljp}) together with
(\ref{pp}) and (\ref{xp}) prescribe
$J_{\mu}$ to be independent of $x_{\mu}$ in both approaches.
This means that in the minimal formulation the transverse part
$J^{\bot}_{\mu}$ of the spin addition $J_\mu$ can depend only on $p_{\mu}$.
Then in both cases the brackets (\ref{caljj}) lead to the condition
\begin{equation}
\{J_{\mu},J_{\nu}\}=-\epsilon_{\mu\nu\lambda}
\left(J^{\lambda}+(p^{\sigma}\partial^{\lambda}-p^{\lambda}\partial^{\sigma})
J_{\sigma}-p^{\lambda}R^{\sigma}p_{\sigma}\right).
\label{jjgen}
\end{equation}
Here and below we use the notation
$\partial^{\mu}=\partial/\partial p_{\mu}$.
Moreover, the functions $R_{\mu}$
defining the brackets of the coordinates must be chosen in such a way,
that the Poisson brackets for independent phase space
variables $x_{\mu}$, $p_{\mu}$ (and $z_{n}$ in the extended case)
would satisfy corresponding Jacobi identities. The
Jacobi identity $\{\{x_{\mu},x_{\nu}\},p_{\lambda}\}+cycle=0$
is reduced to the condition of independence of $R_{\mu}$ on $x_{\nu}$,
whereas the identity
$\{\{x_{\mu},x_{\nu}\},x_{\lambda}\}+cycle=0$
is reduced to the condition
\begin{equation}
\partial^{\mu}R_{\mu}=0.
\label{jacobi}
\end{equation}
Relations (\ref{jjgen}) and (\ref{jacobi}) will play an important role
in what follows.

\section{Minimal canonical approach}
In this section we shall analyse in detail the minimal canonical
approach.  First we shall find the most general admissible form
for the quantities $R_\mu$ defining the brackets of coordinates
of the particle, and, besides, we shall establish
the form  of the transverse part
$J^{\bot}_\mu$ of the angular momentum addition $J_\mu$. Then we
shall analyse the Lorentz properties of the coordinates $x_\mu$,
and determine the form of the angular momentum addition and the
form of the brackets $\{x_\mu,x_\nu\}$ compatible with the
requirement of covariant (vector) behavior of the coordinates
with respect to the Lorentz transformations.  The form of the
brackets will turn out to be a nontrivial one corresponding to the
monopole-like symplectic two-form considered in refs.
\cite{chou1,jn1,ss,chou2}.
After that we shall find the general form for the total angular
momentum vector ${\cal J}_\mu$ compatible with the trivial
brackets $\{x_\mu,x_\nu\}=0$, and shall determine the corresponding
Lorentz properties of $x_\mu$.  In conclusion of this section we
shall comment on the quantum theory corresponding to the minimal
canonical approach.

As we have pointed out in the end of the previous section, in the minimal
approach $J_{\mu}=J_{\mu}(p)$, and, so, $\{J_{\mu},J_{\nu}\}=0$.  Therefore,
the transverse part of $J_\mu$ can be presented as
\begin{equation}
J^{\bot}_{\mu}=-\epsilon_{\mu\nu\lambda}A^{\nu}p^{\lambda},
\label{jbot}
\end{equation}
where $A_{\mu}$ is some function of $p_{\mu}$ being defined up to the
arbitrary term of the form $p_{\mu}\cdot a(p)$ and having the dimensionality
$[A]=[p]^{-1}$. Then we find the condition (\ref{jjgen}) is equivalent to
the condition
$
p^{\mu}(R_{\mu}-\epsilon_{\mu\nu\lambda}\partial^{\nu}A^{\lambda})=
s/\sqrt{-p^{2}},
$
and, therefore,
\begin{equation}
R_{\mu}=-s\frac{p_{\mu}}{(-p^{2})^{3/2}}
+\epsilon_{\mu\nu\lambda}\partial^{\nu}A^{\lambda}
+R^{\bot}_{\mu}
\label{rmu*}
\end{equation}
solves eq. (\ref{jjgen}), where $R^{\bot}_{\mu}$ is an addition orthogonal
to $p_{\mu}$, $R^{\bot}_{\mu}p^{\mu}=0$.
Now, using eq. (\ref{rmu*}), we find that eq. (\ref{jacobi})
is reduced to
the condition $\partial^{\mu}R^{\bot}_{\mu}=0$,
whose general solution is $R^{\bot}_{\mu}=-\epsilon_{\mu\nu\lambda}
p^{\nu}\partial^{\lambda}g$ with an arbitrary function $g=g(p)$,
$[g]=[p]^{-2}$.
Therefore, as a general solution to eqs.
(\ref{jjgen}) and (\ref{jacobi}) we get finally the functions
\begin{equation}
R_{\mu}=-s\frac{p_{\mu}}{(-p^{2})^{3/2}}+\epsilon_{\mu\nu\lambda}\partial^{\nu}
\tilde{A}{}^{\lambda}
\label{rmu1}
\end{equation}
containing arbitrary functions $\tilde{A}_{\mu}=A_{\mu}+p_{\mu}g$.

Hence, a (2+1)-dimensional
relativistic massive particle with arbitrary spin $s$
can be described by the coordinates $x_{\mu}$ and conjugate momenta
$p^{\mu}$ with brackets given by eqs. (\ref{pp}), (\ref{xp})
and (\ref{xxr}), where $R_\mu$ is given, in turn, by eq. (\ref{rmu1}).
Constraint (\ref{p2m2}) prescribes the particle to have a fixed mass $m>0$,
whereas the nontrivial value of spin $s$ is coded in the form of the total
angular momentum vector
\begin{equation}
{\cal J}_{\mu}=-\epsilon_{\mu\nu\lambda}x^{\nu}p^{\lambda}
-se^{(0)}_{\mu} -\epsilon_{\mu\nu\lambda}A^{\nu}p^{\lambda}.
\label{caljm}
\end{equation}
Its specific form together with
nontrivial form of brackets (\ref{xxr}), (\ref{rmu1})
guarantees the fulfillment of the
Poincar\'e algebra (\ref{pp})--(\ref{caljp}).

Here a remark is in order.  The difference between $\tilde{A}_{\mu}$ from eq.
(\ref{rmu1}) and $A_{\mu}$ from eq. (\ref{caljm}) coincides with the
arbitrariness up to which $A_{\mu}$ has been defined itself when we have
introduced it into consideration in eq. (\ref{jbot}).  This difference can be
removed by redefining the coordinates:
$
x_{\mu}\rightarrow x^{\prime}_\mu=x_{\mu}+p_{\mu}g.
$
Such a redefinition does not change the form of the total angular momentum
vector (\ref{caljm}), but the functions $R_{\mu}$ giving the brackets for the
redefined coordinates take the form
\begin{equation}
R_{\mu}=-s\frac{p_{\mu}}{(-p^{2})^{3/2}}
+\epsilon_{\mu\nu\lambda}\partial^{\nu}A^{\lambda}.
\label{rmua}
\end{equation}
So, without loss of generality,
within the minimal approach the brackets for
the coordinates can be taken in the form (\ref{xxr}) with the functions
$R_{\mu}$ defined by eq. (\ref{rmua}).  Further on
we shall use this form of
the functions $R_{\mu}$ as giving the general case.

Let us pass over to the analysis of the Lorentz properties of the
coordinates  $x_\mu$. From the form of the brackets
\begin{equation}
\{{\cal J}_{\mu},x_{\nu}\}=-\epsilon_{\mu\nu\lambda}x^{\lambda}
-\eta_{\mu\nu}\epsilon_{\lambda\rho\sigma}p^{\lambda}\partial^{\rho}
A^\sigma
+\epsilon_{\mu\rho\sigma}(p_{\nu}\partial^{\rho}-\partial_{\nu}p^{\rho})
A^{\sigma}
\label{nolor}
\end{equation}
we conclude that in the general case the three $x_{\mu}$, $\mu=0,1,2$, has
transformation properties different from those of a Lorentz vector.
Simple algebraic calculations show that the coordinates
of the particle $x_\mu=x^c_\mu$
have covariant transformation properties given by the
brackets
\begin{equation}
\{{\cal J}_{\mu},x^{c}_{\nu}\}=-\epsilon_{\mu\nu\lambda}x^{c}{}^{\lambda}
\label{covar}
\end{equation}
only in the case when $A^\mu$ has a special form
$A^{\mu}=p^{\mu}a$, where $a=a(p^{2})$,
$[a]=[p]^{-2}$, is an arbitrary function.
In its turn, this means that the coordinates of the particle have a covariant
nature when
\begin{equation}
{\cal J}_{\mu}=-\epsilon_{\mu\nu\lambda}x^{c}{}^{\nu}p^{\lambda}-
se_{\mu}^{(0)},
\label{caljc}
\end{equation}
and
\begin{equation}
\{x^{c}_{\mu},x^{c}_{\nu}\}=-s\epsilon_{\mu\nu\lambda}
\frac{p^{\lambda}}{(-p^{2})^{3/2}}.
\label{xxc}
\end{equation}
Therefore, the coordinates of the particle have a covariant sense only in the
case when their brackets as well as the total angular momentum vector have
special fixed form, given by eqs. (\ref{xxc}) and (\ref{caljc}),
respectively.  The form of brackets (\ref{xxc}) means that in the quantum
case there is no representation in which the operators corresponding to the
covariant coordinates $x^c_\mu$ would be diagonal. We shall return to this
point in the end of the section.  Another specific feature which we have here
is that only in this covariant case the spin addition
$J^{c}_\mu=
-se^{(0)}_{\mu}$ in the total angular momentum vector (\ref{caljc}) is
parallel to the energy-momentum vector $p_{\mu}$. It is necessary to stress
that due to the nontrivial brackets (\ref{xxc}), neither the first ``orbital"
term from ${\cal J}_{\mu}$ nor the second one do not satisfy
the Lorentz algebra
at the level of Poisson brackets, but only the total vector
(\ref{caljc}) does.  Therefore, we conclude that the information on the
spin of the system is coded simultaneously in the form
of the total angular momentum vector and in the form of the brackets for the
covariant coordinates, and only bearing in mind this fact one can
call the second
term $J^{c}_\mu=-se^{(0)}_\mu$ as the spin vector.  Moreover, we see that the
properties of the covariance (\ref{covar}) and localizability of the
coordinates $x_\mu=x^l_\mu$,
\begin{equation}
\{x^{l}_{\mu},x^{l}_{\nu}\}=0,
\label{local}
\end{equation}
are compatible iff $s=0$. It is necessary also to note that the general
case, given by eqs. (\ref{caljm})  and (\ref{rmua}), is connected
with the special
case (\ref{caljc}), (\ref{xxc}) with covariant coordinates $x^{c}_{\mu}$
through the relation
\begin{equation}
x^{c}_{\mu}=x_{\mu}+A_{\mu}.
\label{transf}
\end{equation}
As we shall see below, this relation turns out to be important
in the construction of the corresponding quantum scheme for the minimal
approach.

Now, let us find the general form for the total angular momentum vector
${\cal J}_{\mu}$ compatible with the property of localizability of the
coordinates (\ref{local}).  In correspondence with eq. (\ref{rmua}),
equality (\ref{local}) takes place when $A_{\mu}$
is subject to the equation
\begin{equation}
\partial_{\mu}A_{\nu}-\partial_{\nu}A_{\mu}=
-s\epsilon_{\mu\nu\lambda}\frac{p^{\lambda}}{(-p^{2})^{3/2}}.
\label{monopol}
\end{equation}
This equation means that the ``gauge field"  $A_\mu$ has the curvature
of the SO(2,1) monopole and is
defined up to the ``gauge transformation"
\begin{equation}
A_{\mu}\rightarrow A'_{\mu}=A_{\mu}+\partial_{\mu}f,
\label{gt}
\end{equation}
where $f=f(p)$ is an arbitrary function of dimensionality $[f]=[p]^0$.
Up to the ``gauge transformation" (\ref{gt}), the
``gauge potential" can be chosen in the form:
\begin{equation}
A_{\mu}=-s\epsilon_{\mu\nu\lambda}\frac{p^{\nu}\xi^{\lambda}}
{p^{2}+\sqrt{-p^{2}}(p\xi)},
\label{del}
\end{equation}
where $\xi^{\mu}$ is an arbitrary fixed timelike unit vector,
$\xi^{2}=-1$.
Since the denominator on the r.h.s. of eq. (\ref{del}) contains
the factor $(1-e^{(0)}\xi)$,
we can choose the vector $\xi_{\mu}$ in the form
\begin{equation}
\xi^{\mu}=\frac{p^{0}}{\vert p^{0}\vert}\cdot (1,0,0)
\label{xif}
\end{equation}
in order to avoid a singularity not only for $p^{0}>0$ but also
for $p^{0}<0$.
Such a choice of $\xi^\mu$ is possible since from the very beginning we have
borne in mind that $p^{2}<0$. Hence, the sectors with $p^{0}>0$ and $p^{0}<0$
are separated (i.e. the phase space is supposed to be disconnected), and
therefore, the sign factor in eq. (\ref{xif}) can be considered as a constant
one in these two sectors.
On the other hand, if we omit the sign factor in eq. (\ref{xif}), we
will have the singularity for the case $p^0<0$:  $A_{i}\propto
\epsilon^{ij}p^{j}/p^{k}p^{k}$ as $p^{i}\rightarrow 0$, $i=1,2$,
where we have introduced the notation $\epsilon^{ij}=\epsilon^{0ij}$.

Let us note here that due to the form of the total angular momentum vector
(\ref{caljm}), the gauge
transformation (\ref{gt}) can be ``absorbed" into the first, orbital-like
term by redefining the coordinates, $x_{\mu}\rightarrow x_{\mu}+
\partial_{\mu}f$, since such a redefinition is a canonical
transformation. Therefore, without loss of generality we can use the
fixed form of the ``gauge potential" (\ref{del}) with $\xi_{\mu}$ given by eq.
(\ref{xif}).

Thus, the specific form of the total angular momentum (\ref{caljm})
with the SO(2,1) ``monopole gauge potential" $A_{\mu}$
given by eq. (\ref{del})
guarantees the fulfillment of algebra (\ref{caljj})
in the case when the coordinates of the
particle have the property of localizability (\ref{local}).
But as we have pointed out above,
in this case the coordinates of the particle
have noncovariant properties
given by eq. (\ref{nolor}), where $A_{\mu}$, in turn, is given
by eqs. (\ref{del}), (\ref{xif}).
Indeed, in the case under consideration the total angular momentum
is presented in the form
\begin{equation}
{\cal J}_{\mu}=-\epsilon_{\mu\nu\lambda}x^{l\nu}p^{\lambda}+J^{l}_{\mu},
\label{jex}
\end{equation}
where
\begin{equation}
J^{l}_0=-s\frac{p_{0}}{\vert p^{0}\vert },\qquad
J^{l}_i=-s\frac{p_{i}}{\sqrt{-p^{2}}+\vert p^0\vert},
\label{jex1}
\end{equation}
and we have the transformation properties of the coordinates $x^{l}_{\mu}$
given by the brackets
\begin{equation}
\{{\cal J}_{\mu},x^{l}_{\nu}\}=-\epsilon_{\mu\nu\lambda}x^{l\lambda}
+\Delta_{\mu\nu},
\label{xll}
\end{equation}
where
\[
\Delta_{0\nu}=0,\qquad
\Delta_{i\nu}=\frac{s}{\vert p^{0}\vert +\sqrt{-p^{2}}}
\cdot e^{(i)}_{\nu},
\]
with
\begin{equation}
e^{(i)0}=\frac{p^{0}}{\vert p^{0}\vert}\cdot \frac{p^{i}}{\sqrt{-p^{2}}},
\quad
e^{(i)j}=\delta^{ij}+\frac{p^{i}p^{j}}{\sqrt{-p^{2}}(\vert p^{0}\vert
+\sqrt{-p^{2}})}.
\label{eij}
\end{equation}
The presence of the noncovariant term $\Delta_{\mu\nu}$ in eq. (\ref{xll})
means that the coordinates $x^{l}_{\mu}$ have noncovariant transformation
properties with respect to the Lorentz boosts, whose generators are ${\cal
J}_{i}$.  The quantities $e^{(i)}_{\mu}$ appearing here are the components of
the triad (\ref{triad}) satisfying
another orientation condition:
$\epsilon_{\mu\nu\lambda}e^{(0)\mu}e^{(1)\nu}e^{(2)\lambda}= -{p^{0}}/{\vert
p^{0}\vert}$.

Let us stress once more that though $J^{l}p
=s\sqrt{-p^2}$, nevertheless, due to
the relation $J^{l}e^{(i)}\neq 0$,
here the spin addition $J^{l}_\mu$ defined by eq.
(\ref{jex1}) is not parallel to the energy-momentum vector $p_\mu$ in
correspondence with the general results declared after eq.  (\ref{xxc}).

Due to noncovariant properties of the localizable coordinates $x^l_\mu$, one
could forget about them and work only in terms of the covariant coordinates
$x^c_\mu$. But we are forced to recall about localizable coordinates as soon
as we pass over to the consideration of the quantum theory. Indeed, due to
nontrivial form of brackets (\ref{xxc}), we have a problem of constructing
the operators corresponding to the classical coordinates $x^c_\mu$.  This
problem can be solved with the help of the localizable coordinates $x^l_\mu$,
which we redenote here as $q_\mu$, in the following way.  According to eq.
(\ref{transf}), we can realize the operators $\hat{x}{}^c_\mu$ with the help
of equality
\begin{equation}
\hat{x}{}^c_\mu=\hat{q}_\mu+\hat{A}_\mu,
\label{opxc}
\end{equation}
where $\hat{A}_\mu$ is the quantum analog of the ``gauge potential" given by
eqs. (\ref{del}) and (\ref{xif}).  Therefore, we can choose the momentum
representation with the operators $\hat{p}_\mu$ being diagonal, and in
correspondence with classical relations (\ref{xp}) and (\ref{local}), the
operators $\hat{q}_\mu$ will be realized as $\hat{q}_\mu=
i\partial/\partial p^\mu$.
On the other hand, we can choose the coordinate representation with the
operators $\hat{q}_\mu$ being diagonal and with
$\hat{p}_\mu=-i\partial/\partial q^\mu$. In this case due to the specific
form of the ``gauge potential" $A_\mu$ being present on the r.h.s. of
eq. (\ref{opxc}), the operator $\hat{x}{}^c_\mu$ will have a
nonlocal form.  In both cases of the momentum and coordinate
representations, noncovariant properties of the operator
$\hat{q}_\mu$ are coded in the nontrivial form of the angular
momentum operator \[ \hat{\cal J}_\mu=-\epsilon_{\mu\nu\lambda}
\hat{q}{}^\nu\hat{p}{}^{\lambda}+\hat{J}{}^{l}_\mu,
\]
where $\hat{J}{}^l_\mu$ is the
quantum analog of the spin addition given by eq.
(\ref{jex1}). This operator being rewritten in terms of
$\hat{x}{}^c_\mu$ with
the help of relation (\ref{opxc}) takes, clearly, the quantum form
corresponding to the classical equality (\ref{caljc}).

Therefore, we conclude that the manifest covariance of the minimal
approach being formulated in terms of the coordinates
$x^c_\mu$ is inevitably lost under transition to the quantum theory,
where we are forced to use localizable, $[\hat{q}_\mu,
\hat{q}_\nu]=0$, but noncovariant operators $\hat{q}_\mu$.
Moreover, in the coordinate representation with diagonal operators
$\hat{q}_\mu$,
the covariant operators $\hat{x}{}^c_\mu$ turn out to be
nonlocal operators. These shortcomings can be removed by using
the extended canonical formulation for the relativistic
fractional spin particles.

\section{Extended canonical formulation}

Let us turn to the general analysis of the extended formulation and its
relation to the minimal one.
As we have pointed out in sect. 2, in the extended case the spin addition
$J_{\mu}$ in the total angular momentum vector ${\cal J}_\mu$ depends on
auxiliary internal variables $z_{n}$ and generally it can also depend on
$p_{\mu}$ (see eq. (\ref{jjgen})).
It is reasonable to restrict this general analysis by
adding the rather natural assumption that $J_{\mu}$ depends only on the
internal variables, $J_{\mu}=J_{\mu}(z_{n})$.  Moreover, let us suppose that
the coordinates of the particle $x_\mu$ are localizable, i.e. they
have the brackets
\begin{equation}
\{x_{\mu},x_{\nu}\}=0.
\label{cloc}
\end{equation}
As a result, the general condition (\ref{jjgen}) takes
the very simple form
\begin{equation}
\{J_{\mu},J_{\nu}\}=-\epsilon_{\mu\nu\lambda}J^{\lambda}.
\label{jj}
\end{equation}
Therefore, under the assumptions made above,
the components of
$J_{\mu}$ themselves must form the Lorentz algebra $so(2,1)$, and
in this case both $J_{\mu}$ and $x_{\mu}$ are Lorentz vectors,
i.e. here the property of the covariance of the coordinates is
compatible with that of localizability (\ref{cloc}) thanks
to the introduction of the auxiliary internal variables.

Now, let us reveal the relation of this formulation to the minimal one,
whereas concrete examples of models realizing the extended formulation
will be considered in the two subsequent sections.  To this end, we decompose
the vector $J_{\mu}=J_{\mu}(z_{n})$ into longitudinal and transverse parts:
\begin{equation}
J_{\mu}=-J^{(0)}e^{(0)}_{\mu}+J^{\bot}_{\mu},
\label{decomp}
\end{equation}
where $J^{(0)}=J^{\mu}e^{(0)}_{\mu}$, $J^{\bot}_{\mu}=(\eta_{\mu\nu}+
e^{(0)}_{\mu}e^{(0)}_{\nu})J^{\nu}$,
and taking into account the mass shell constraint (\ref{p2m2}),
we present the spin constraint (\ref{pjsm}) in the equivalent
form
\begin{equation}
J^{(0)}-s\approx 0.
\label{j0s}
\end{equation}
All the quantities in the system can be classified as those being
either gauge-invariant or gauge-noninvariant ones with respect to the gauge
transformations generated by this constraint \cite{sunder}.
In particular, the brackets
\begin{equation}
\{J^{(0)},x_{\mu}\}=-\frac{1}{\sqrt{-p^{2}}}J^{\bot}_{\mu}
\label{j0x}
\end{equation}
mean that the coordinates $x_{\mu}$ are not gauge-invariant
quantities.

With the help of decomposition  (\ref{decomp}), we can present the total
angular momentum (\ref{calj}) in the equivalent form
\begin{equation}
{\cal J}_{\mu}=-\epsilon_{\mu\nu\lambda}\tilde{x}{}^{\nu}p^{\lambda}-
J^{(0)}e^{(0)}_{\mu}
\label{caljtil}
\end{equation}
with
\begin{equation}
\tilde{x}_{\mu}=x_{\mu}+\frac{1}{p^{2}}\epsilon_{\mu\nu\lambda}p^{\nu}
J^{\lambda}.
\label{tildx}
\end{equation}
The redefined coordinates $\tilde{x}_{\mu}$ form a Lorentz vector
similar to the initial coordinates $x_{\mu}$, but unlike the latter ones,
they have the trivial brackets with $J^{(0)}$,
\begin{equation}
\{\tilde{x}_{\mu},J^{(0)}\}=0.
\label{txj0}
\end{equation}
Hence, $\tilde{x}_\mu$
is the gauge-invariant extension \cite{sunder} of the initial vector $x_\mu$.
The total angular momentum vector (\ref{caljtil}) written in terms
of the gauge-invariant coordinates (\ref{tildx}) has the same form
(\ref{caljc})
(in the weak sense, under taking into account the spin constraint
(\ref{j0s})) as the total
angular momentum vector from the minimal canonical
formulation had when it was
presented in terms of the covariant coordinates $x^c_\mu$.
It is necessary to note that the both terms in eq. (\ref{caljtil})
are gauge-invariant vectors.
The brackets for the redefined coordinates,
\begin{equation}
\{\tilde{x}_{\mu},\tilde{x}_{\nu}\}=-J^{(0)}\epsilon_{\mu\nu\lambda}
\frac{p^{\lambda}}{(-p^{2})^{3/2}},
\label{txtx}
\end{equation}
have the same form (in the weak sense) as the covariant coordinates
$x^{c}_{\mu}$ had.  The gauge-invariant term $-J^{(0)}e^{(0)}_\mu$ in
the angular momentum vector (\ref{caljtil}) carries alongside
with the constraint
(\ref{j0s}) and brackets (\ref{txtx}) the information on the nontrivial value
of the spin, and, therefore, we arrive at a complete correspondence with the
special case of the minimal formulation given in terms of the coordinates
$x^c_\mu$.

Let us demonstrate now how the most general case from the minimal
formulation, given by eqs. (\ref{caljm}), (\ref{rmua}), is related to the
extended formulation.  This will also allow us to get some relations
useful for subsequent considerations.

First of all, from the brackets
\begin{equation}
\{J^{(0)},J^{2}\}=0
\label{j0j2}
\end{equation}
 and
\begin{equation}
\{J^{(0)},J^{\bot}_{\mu}\}=\epsilon_{\mu\nu\lambda}e^{(0)\nu}J^{\bot\lambda},
\label{j0t}
\end{equation}
it follows that the constraint (\ref{j0s}) generates the rotations
of the two-dimensional vector ${\bm J}$
with components $J^{(i)}=e^{(i)}_{\mu}J^{\mu}$, $i=1,2$,
where $e^{(i)}_{\mu}$ are the components of the triad
(\ref{triad}).
Therefore, generally
the gauge orbits generated by the action of the constraint
(\ref{j0s}) are one-dimensional spheres
$S^{1}$ in the three-dimensional space with coordinates
$J^{(\alpha)}$, $\alpha=0,1,2$. These orbits are
the sections of the set of two-dimensional hyperboloids
$J^2=-J^{(0)2}+J^{(i)2}=const$
by the hyperplane $J^{(0)}=s$.

We can pass over from the initial set of auxiliary
internal coordinates
$z_{n}$, $n=1,\ldots, 2N$, to the set of variables
$J^{(0)},$ $\varphi$ and $\tilde{z}_{a}$, $a=1,\ldots,2(N-1)$, such that
$0<\varphi\leq 2\pi$,
\begin{equation}
\{\varphi,J^{(0)}\}=1,
\label{vj1}
\end{equation}
and
$\{\tilde{z}_{a},J^{(0)}\}=\{\tilde{z}_{a},
\varphi\}=\{\tilde{z}_{a},\tilde{x}_{\mu}\}=
\{\tilde{z}_{a},p_{\mu}\}=0.$ The concrete form of $\tilde{z}_a=
\tilde{z}_{a}(z_n,p^\mu)$ is not important for us here.
The variables $J^{(0)}$ and $\varphi$ form a pair of
action-angle type variables.
As a result, we arrive at the
parametrization for the components $J^{(i)}$ of the two-dimensional
vector ${\bm J}$:
\begin{equation}
J^{(i)}=r\cdot n^i,\quad
r=\sqrt{J^{(0)2}+C(\tilde{z}_{a})},\quad
n^{i}=(\sin\varphi,\cos\varphi).
\label{jip}
\end{equation}
We assume here that $C(\tilde{z}_{a})$ and corresponding
region of possible values of $J^{(0)}$ are defined
by concrete model.
Definition (\ref{tildx}) and eq. (\ref{jip}) lead to the brackets
\begin{equation}
\{\tilde{x}_{\mu},\varphi\}=-e^{(1)\nu}\partial_{\mu}e^{(2)}_{\nu}.
\label{txf}
\end{equation}
The angle variable $\varphi$, obviously, parametrizes the points on the
above-mentioned gauge orbits $S^1$'s.
Therefore, proceeding from the initial set
of the phase space variables $x_{\mu}$, $p_{\mu}$ and  $z_{n}$, we can pass
over to the set of the variables $\tilde{x}_{\mu}$, $p_{\mu}$, $J^{(0)}$,
$\tilde{z}_{a}$ and $\varphi$.
The variables $\tilde{x}_\mu$, $p_\mu$, $J^{(0)}$ and
$\tilde{z}_a$
form the complete set of the variables being gauge-invariant with
respect to the gauge transformations generated by the spin constraint
(\ref{j0s}), whereas $\varphi$ is the only gauge-noninvariant variable.  For
completeness we stress once more
that in accordance with the general discussion from
section 2, it is necessary to supplement the spin constraint (\ref{j0s})  and
the mass shell constraint (\ref{p2m2}) with a
corresponding number of first
and/or second class constraints which would ``freeze" internal
degrees of freedom described by the variables
$\tilde{z}_a$.

Now, it is obvious that the reduction of the system to the surface
of the constraint (\ref{j0s}) can be realized by introducing a
(local) gauge condition
\begin{equation}
\varphi-\varphi_{0}\approx 0,
\label{f0}
\end{equation}
where $\varphi_{0}$ is some fixed point on a
one-dimensional gauge orbit $S^1$.
The Dirac brackets must be
calculated with the help of the pair of second
class constraints (\ref{j0s}) and (\ref{f0}).
Due to the gauge invariance of the quantities $\tilde{x}_{\mu}$, $p_{\mu}$
and ${\cal J}_{\mu}$
with respect to the gauge
transformations generated by the constraint (\ref{j0s}),
all the Dirac brackets between them will coincide
with the corresponding initial brackets. At the same time, we find
that the Dirac brackets $\{x_{\mu},x_{\nu}\}^{*}$ have the form given
by eqs.  (\ref{xxr}) and (\ref{rmua}) with the functions
$A_{\mu}$ given by
\begin{equation}
A_{\mu}=\frac{1}{p^{2}}\epsilon_{\mu\nu\lambda}p^{\nu}J^{\lambda},
\quad J_{\mu}=-se^{(0)}_{\mu}+e^{(i)}_{\mu}\gamma^{(i)},
\label{aj}
\end{equation}
where quantities $\gamma^{i}$ are defined, in turn,  by eq.
(\ref{jip}) at $J^{(0)}=s$ and $\varphi=\varphi_{0}$.

Therefore, we conclude that the gauge-invariant coordinates
(\ref{tildx}) have the same sense here
as the covariant ``nonlocalizable"
coordinates $x^{c}_{\mu}$ from the minimal canonical formulation.
On the other hand, the initial gauge-noninvariant variables
$x_{\mu}$ in the following sense are analogous to the coordinates
corresponding to the general case of the minimal formulation,
given by the total angular momentum vector (\ref{caljm})
and by the functions $R_{\mu}$ (\ref{rmua}):
here the Dirac brackets $\{x_{\mu},x_{\nu}\}^{*}$,
appearing after reduction of the system to the surface defined by
eqs. (\ref{j0s}) and (\ref{f0}), correspond to the brackets
of the coordinates $x_\mu$ from the minimal formulation.

It is necessary to note that after reduction of the system,
in correspondence with the general results obtained for the
minimal formulation, the initial gauge-noninvariant
coordinates $x_{\mu}$ will not generally have covariant
properties with respect to the Lorentz transformations due to a
noncovariant character of eq.  (\ref{f0}) giving the reduced
subspace.  Only in one case, when $-J^{2}=s^{2}$, and, so,
$\gamma^{(i)}=0$ and $A_{\mu}=0$ in eq.  (\ref{aj}),
gauge-invariant coordinates $\tilde{x}_{\mu}$ coincide with
$x_{\mu}$.  From the point of view of the reduction, such a
case can be understood as a limit one, in which the above-mentioned
gauge orbits shrink into just a point. We shall
consider the concrete model comprising such a limit case in the
next section, where, in particular, we shall demonstrate that
this case will also be a special one from the lagrangian point
of view.

Thus, we have revealed the relation of the extended canonical formulation to
the  minimal one through the reduction of the former formulation.
But as we have
pointed out in the beginning of the section, the advantage of the former
formulation is that it allows us to describe the system in terms of the
covariant coordinates $x_\mu$ having commuting components in the sense of the
Poisson brackets (\ref{cloc}) when fixing the spin is realized in the theory
in the weak sense
via the constraint (\ref{pjsm}). So, we pass over to the consideration of
concrete models withing the framework of the extended formulation.

\section{Minimal extended model}

In this section we shall consider the extended model
with a minimal number of
$2N=2$ internal auxiliary phase space variables. At the quantum level,
as we shall see,
this model will correspond to the approach with infinite-dimensional
representations of ${\rm \overline{SL(2,R)}}$ group \cite{pl5,jn1}.

The minimal extended model can be constructed in the following way.
First we note that
the scalar $J^{2}$ lies in the center of the algebra
(\ref{jj}), $\{J^{2},J_{\mu}\}=0$.
Therefore, it can be fixed by introducing the condition
\begin{equation}
J^{2}=C,
\label{const}
\end{equation}
where $C$ is some real constant which corresponds here to
$C(\tilde{z}_{a})$ from
eq. (\ref{jip}). So, if we consider the dependent
variables $J_{\mu}$, subject to the condition
(\ref{const}), as the internal variables themselves, we shall have
the internal phase subspace with $2N=2$ independent variables.
This subspace has the structure defined by the
constant $C$:
for $C=-\alpha^{2}$, $\alpha>0$, eq. (\ref{const}) sets two
disconnected sheets of the hyperboloid:
\begin{equation}
J_{0}=\varepsilon \sqrt{\alpha^{2}+J_{i}^{2}}, \quad \varepsilon=\pm, \quad
i=1,2,
\label{2hyp}
\end{equation}
whereas in the case $C=\beta^{2}$, $\beta>0$,
it defines a one-sheet hyperboloid
degenerating into the cone at $\beta=0$.

Thus, the minimal extended model can be
given by the variables $x_{\mu}$ and $p_{\mu}$
with canonical brackets (\ref{pp}),
(\ref{xp}), (\ref{cloc}), and by the spin variables $J_{\mu}$
subject to the condition (\ref{const}) and forming the
$so(2,1)$ algebra (\ref{jj}). Moreover, the model has
the set of two first
class constraints (\ref{p2m2}) and (\ref{pjsm}). It is necessary
to stress that for the case $J^{2}=-\alpha^{2}<0$,
the spin constraint has nontrivial solutions only when $s^{2}\geq \alpha^{2}$,
and, therefore, the model is defined only for $s^{2}\geq -J^{2}$.

Let us construct now the lagrangian corresponding to the model.
The brackets (\ref{jj}) for the internal variables $J_{\mu}$ can be
derived from a kinetic lagrangian
\begin{equation}
L_{kin}=-\frac{J\zeta}{J^{2}+(J\zeta)^{2}}\epsilon_{\mu\nu\lambda}
\zeta^{\mu}J^{\nu}\dot{J}{}^{\lambda}
\label{lkin}
\end{equation}
containing an arbitrary fixed unit timelike vector $\zeta^{\mu}$,
$\zeta^{2}=-1$.
The simplest way to be convinced that it is  so consists in checking the
fact that under a Lorentz transformation of $J_{\mu}$, the kinetic term
(\ref{lkin}) is changed by a total derivative, and, therefore, it
corresponds to a Lorentz invariant term in the action.
Then, choosing $\zeta^{\mu}=(1,0,0)$, and
parametrizing the variables $J_{\mu}$ as
\begin{equation}
J_{\mu}=\left(J_{0}, \sqrt{J_{0}^{2}+C}\cdot n_{i}\right),\quad
n_{i}=(\cos \varphi,\sin \varphi),
\label{parjmu}
\end{equation}
where $0\leq\varphi<2\pi$ and
$-\infty<J_{0}<\infty$ in the case $C=\beta^2\geq 0$,  whereas
$J_{0}$ can take values in the region $[\alpha,+\infty)$
or $(-\infty,-\alpha]$ when $J^{2}=-\alpha^{2}<0$,
one gets
\[
 L_{kin}=J_{0}\cdot \dot{\varphi}.
\]
{}From here we find that the brackets for independent variables
$J_{0}$ and $\varphi$ have the form
\begin{equation}
\{\varphi,J_{0}\}=1.
\label{phij}
\end{equation}
Using these brackets and parametrization (\ref{parjmu}),
one can get convinced that
the lagrangian (\ref{lkin}) indeed leads to the brackets (\ref{jj}).

The total lagrangian of the model can be obtained by the inverse
Legendre transformation
from the requirement that it would lead
to the constraints (\ref{p2m2}) and (\ref{pjsm}). This gives
\begin{equation}
L=\frac{1}{2e}(\dot{x}_{\mu}-vJ_{\mu})^{2}-\frac{1}{2}em^{2}+
s m v+L_{kin}.
\label{ltot}
\end{equation}
Lagrangian (\ref{ltot}), with $e$ and $v$ being the  Lagrange multipliers,
leads to the primary constraints
\begin{equation}
p_{e}\approx 0,\quad p_{v}\approx 0,
\label{pepv}
\end{equation}
and to the mass shell and spin conditions (\ref{p2m2}) and (\ref{pjsm})
as the secondary constraints,
where $p_e$ and $p_v$ are the momenta conjugate to $e$ and $v$,
$\{e,p_e\}=1$, $\{v,p_v\}=1$.

The corresponding action $A=\int Ld\tau$
is invariant with respect to the reparametrizations:
\begin{eqnarray}
&\delta x_{\mu}=\gamma \dot{x}_{\mu},\quad
\delta J_{\mu}=\gamma \dot{J}_{\mu},\quad
\delta e=\frac{d}{d\tau}(\gamma e),\quad
\delta v=\frac{d}{d\tau}(\gamma v),&\nonumber\\
&\delta L=\frac{d}{d\tau}(\gamma L),&
\label{gauge1}
\end{eqnarray}
whose generator is the mass shell constraint,
and with respect to the transformations
generated by the spin constraint (\ref{pjsm}):
\begin{eqnarray}
&\delta e=0,\quad \delta v=\dot{\rho},\quad
\delta x_{\mu}=\rho J_{\mu},\quad
\delta J_{\mu}=-\rho e^{-1}\epsilon_{\mu\nu\lambda}\dot{x}{}^{\nu}J^{\lambda},
&\nonumber\\
&\delta L=\frac{d}{d\tau}\left(\rho\left(sm+
e^{-1}\dot{x}J -J^{2}e^{-1}\left(\dot{x}J+(\dot{x}\zeta)(\zeta J)\right)
\cdot(J^{2}+(J\zeta)^{2})^{-1}\right)\right).&
\label{gauge2}
\end{eqnarray}
Here $\gamma=\gamma(\tau)$ and $\rho=\rho(\tau)$
are infinitesimal parameters of the transformations.

Let us consider now the Lagrange equations of motion for
$e$ and $v$,
\begin{equation}
(\dot{x}_{\mu}-vJ_{\mu})^{2}+e^{2}m^{2}=0,\quad
\dot{x}J-vJ^{2}-s me=0.
\label{lcon}
\end{equation}
{}From the second equation we get the equality
$
e=s^{-1}m^{-1}(\dot{x}J-vJ^{2}),
$
and we can rewrite the first equation in the form
\[
\dot{x}^{2}+s^{-2}(\dot{x}J)^{2}=-v(1+s^{-2}J^{2})
\cdot (vJ^{2}-2\dot{x}J).
\]
{}From here we conclude that iff
\begin{equation}
-J^{2}=\alpha^{2}=s^{2},
\label{scase}
\end{equation}
there is the Lagrange constraint:
\begin{equation}
\dot{x}^{2}+s^{-2}(\dot{x}J)^{2}=\dot{x}^{2}-(\dot{x}J)^{2}\cdot
(J^{2})^{-1}=0,
\label{lagc}
\end{equation}
which means that the particle velocity vector $\dot{x}_{\mu}$ is parallel
to the spin vector $J_{\mu}$.
Note, that the analogous property of parallelness between $p_{\mu}$ and
$J_{\mu}$ in the case (\ref{scase}) takes place in this model
at the hamiltonian level. It is important to stress here
that such a property of parallelness of the spin vector
$J_\mu$ to $p_\mu$ is valid only in the weak sense,
on the surface of the spin constraint. Therefore, in the case (\ref{scase})
the particle velocity turns out also to be parallel to its energy-momentum
vector $p_\mu$. Below we shall return to this point.

One can rewrite lagrangian (\ref{ltot}) in a form
revealing the speciality of the case (\ref{scase}) in a more explicit
way. To this end, we find the multiplier $v$ from the second
equation (\ref{lcon}) assuming that $J^{2}\neq 0$. We get
$v=(J^{2})^{-1}(\dot{x}J-sme)$,
and after substituting this value into lagrangian (\ref{ltot}),
we arrive at the following form for the total lagrangian:
\begin{equation}
L=\frac{1}{2e}\left(\dot{x}{}^{2}-(J^{2})^{-1}
(\dot{x}J)^{2}\right)+sm(J^{2})^{-1}(\dot{x}J)
-\frac{1}{2}em^{2}\left(1+s^{2}(J^{2})^{-1}\right)+L_{kin}.
\label{lspe}
\end{equation}
The term linear in $e$ disappears from (\ref{lspe}) only
when eq. (\ref{scase}) takes place. Therefore, in this case
the variation of the corresponding action over $e$
gives directly the Lagrange constraint (\ref{lagc}).
The spin constraint (\ref{pjsm})
appears from lagrangian (\ref{lspe})
for this special case as the primary constraint,
whereas the mass-shell constraint (\ref{p2m2}) is a secondary one.

Let us note that the general procedure of the
reduction of the system, described in the
previous section, can be applied also to the special case (\ref{scase})
if to consider it as a limit, e.g., as $J^2=-\alpha^2$,
$s^2=\alpha^2+\epsilon^2$, $\epsilon\rightarrow 0$.
After such a reduction we get the strong equality $J_\mu=-se^{(0)}_\mu$.
Due to the gauge-noninvariance of the vector $J_\mu$
(see eq. (\ref{j0t})), instead of the initial $so(2,1)$ algebra
(\ref{jj}) we arrive as a result of the reduction
at the trivial Dirac brackets for the components
of the vector $J_\mu$,
which turns out to be parallel to the energy-momentum vector $p_\mu$
in the strong sense on the reduced phase space.

Let us investigate now the classical motion of the particle in the model
under consideration.  It can be done in the simplest way
within the framework
of the hamiltonian approach proceeding from the initial form of lagrangian
(\ref{ltot}). The corresponding total hamiltonian has here the form
\begin{equation}
H=\frac{e}{2}(p^2+m^2)+v(pJ-ms)+w_1 p_e +w_2 p_v,
\label{hamtot}
\end{equation}
where $w_{1,2}=w_{1,2}(\tau)$
are arbitrary functions of the evolution parameter, which are
associated with the primary first class constraints (\ref{pepv}) \cite{sunder}.
The equations of motion generated by the total hamiltonian via
the Poisson brackets have the form:
\begin{eqnarray}
&\dot{p}_\mu=0,\quad
\dot{x}_\mu=ep_\mu+vJ_\mu,\quad
\dot{J}_\mu=-v\epsilon_{\mu\nu\lambda}p^\nu J^\lambda ,&\label{eq1}
\\
&\dot{e}=w_1,\quad
\dot{v}=w_2 .&
\label{eq2}
\end{eqnarray}
{}From here we see once again that since generally $J_\mu$ is not parallel
to the conserved energy-momentum vector $p_\mu$, the velocity $\dot{x}_\mu$
is not parallel to $J_\mu$.
Note, that the coordinates $\tilde{x}_\mu$,
constructed according to eq. (\ref{tildx}), have the evolution
law given by the equation
\begin{equation}
\dot{\tilde{x}}_\mu=(e-vsm^{-1})\cdot p_\mu,
\label{eq3}
\end{equation}
where we have taken into account the mass shell and spin constraints.
As we shall see below, the coordinates of the particle $x_\mu$, being
gauge-noninvariant quantities with respect to the gauge transformations
generated by the constraint (\ref{j0s}), have the evolution law
revealing, unlike the gauge-invariant coordinates $\tilde{x}_\mu$,
the classical analog of the relativistic quantum
{\it Zitterbewegung} which
takes place in the general case for models of spinning particles \cite{pl2}.

Contracting $x_\mu$ and $J_\mu$ with the triad components
$e^{(\alpha)}_\mu$ given by eq. (\ref{eij})
and taking again into account the mass shell and spin constraints,
we get
\begin{eqnarray}
&\dot{x}{}^{(0)}=-me+vs,\quad \dot{x}{}^{(i)}=vJ^{(i)},&
\label{eq4}\\
&J^{(0)}=s,\quad \dot{J}{}^{(i)}=
-\varepsilon^0 mv\epsilon^{ij}J^{(j)},&
\label{eq5}
\end{eqnarray}
where through $\varepsilon^0$ we have denoted the sign of the energy $p^{0}$.
Now let us fix the variables $v$ and $e$ as
\begin{equation}
v-v_0=0,\quad
e-m^{-1}(1+v_0 s)=0,
\label{vegau}
\end{equation}
where $v_0$ is a constant, restricted by the condition
\begin{equation}
v_0^2\leq r^{-2},\quad r=\sqrt{C+s^2},
\label{restrv0}
\end{equation}
(see below).
Conditions (\ref{vegau})
are really the gauge conditions conjugate to the primary
constraints (\ref{pepv}), and the requirement of their stationarity leads,
according to eq. (\ref{eq2}), to fixing the multipliers
$w_{1,2}$ in the total
hamiltonian: $w_1=w_2=0$.
{}From the lagrangian point of view the introduction of conditions
(\ref{vegau}) corresponds, obviously, to fixing the gauge freedom
given by eqs. (\ref{gauge1}) and (\ref{gauge2}).
Such a choice leads to a simple
form for the general solutions to eqs. (\ref{eq4}) and (\ref{eq5}):
\begin{equation}
x^{(0)}=\varepsilon^{0}(\tau+\tau_0),\quad
x^{(i)}=m^{-1}rk^{(i)}+x^{(i)}_0,\quad
J^{(i)}=rn^{(i)}.
\label{eq6}
\end{equation}
Here $k^{(1)}=n^{(2)}= \sin\varphi(\tau),$
$k^{(2)}=n^{(1)}=\cos\varphi(\tau)$,
$\varphi(\tau)=-\varepsilon^0 mv_0\tau+\varphi_0$,
and $\tau_0$, $x^{(i)}_0$ and $\varphi_0$  are some constants of
integration. The first equality means that in the rest frame system,
$p^i=0$, $i=1,2$, we have in fact the laboratory temporal gauge:
$x^0=\varepsilon^0 (\tau+\tau_0)$.
According to the second relation and eq. (\ref{eij}) defining the explicit
form of the triad components $e^{(i)}_\mu$,
in this system the particle performs a circular motion.
The radius of the circle, $m^{-1}r$, is defined by the choice of the
model's constant $C$, whereas the velocity is a gauge-dependent quantity,
$\dot{x}{}^{(i)}\dot{x}{}^{(i)}=v_0^2 r^2\leq 1$.
In the case of arbitrary momentum ${\bm p}\neq {\bf 0}$ we have here a
relativistic superposition of the circular motion and the
rectilinear motion with a constant
velocity (in the laboratory time) equal to ${\bm p}/p^0$.
Indeed, as follows from eqs. (\ref{eq3})
and (\ref{vegau}), here
\begin{equation}
\dot{\tilde{x}}{}^0=\frac{p^0}{m}
\label{tx0}
\end{equation}
and $\dot{\tilde{x}}^i=p^i/m$. Therefore,
the gauge-invariant coordinates $\tilde{x}_\mu$
do not reveal any gauge-dependent circular motion, and, moreover,
\begin{equation}
\frac{d\tilde{x}^i}{d\tilde{x}^0}=\frac{p^i}{p^0}
\label{foldi}
\end{equation}
in a complete correspondence with the ordinary
evolution law for relativistic free particle.
Using eqs. (\ref{tx0}) and (\ref{foldi}),
the definition of the gauge-invariant
coordinates (\ref{tildx}) and the form of the triad
(\ref{eij}), one can get convinced that in the general case, when
${\bm p}\neq {\bf 0}$, the coordinates $x^i$ evolves
in the way that has been described above.

Here it is necessary to make an important remark.  Using eq. (\ref{tx0}) and
definition (\ref{tildx}), one can find that the inequality $\dot{x}{}^0\neq
0$ takes place only in the case when the velocity of the particle is not
greater
than the velocity of light.  It means, in turn, that
only when
$\dot{x}{}^2\leq 0$,
one can pass over to the
laboratory temporal gauge with $\vert\dot{x}{}^0\vert=1$ for any value of the
momentum ${\bm p}$, and, as consequence, the coordinate $x^0$ can be
interpreted as a time only in this case.  It is due to
this reason we have introduced the restriction on the modulus of the constant
$v_0$ in eq. (\ref{restrv0}) (for a more detailed discussion of this
point see ref. \cite{pl2}).

Thus, we conclude that generally the coordinates of the particle
$x_i$ are subject to the gauge-dependent circular motion being
the classical analog of the relativistic quantum
Zitterbewegung \cite{zit}.
Therefore, the equality to zero
of the second time derivative of the coordinates of the particle
is not a necessary condition for the relativistic spinning
particle to be a free particle in 2+1 dimensions.  Only in one
case, when $C=J^2=-s^2$, there is no such a circular motion in
the system and here $\ddot{x}_\mu=0$.  At the same, the space
components of the gauge-invariant coordinate $\tilde{x}_\mu$ do
not reveal a Zitterbewegung and evolves according to the
ordinary law for the coordinates of the free relativistic
spinless particle.  Note, that in this respect the latter
coordinates are analogous to the Foldy-Wouthuysen coordinates
of the Dirac particle \cite{foldy}.

Now let us pass over to the discussion of the quantization of the model.
We have here covariant variables $x_\mu$
with classical brackets (\ref{cloc}), and, so, can choose the coordinate
representation with the operators $\hat{x}_\mu$ being diagonal and
with $\hat{p}_\mu=-i\partial/\partial x^\mu$. Then the only problem
consists in realizing the quantum analogs of the variables $J_\mu$
subject to the condition (\ref{const}) and forming with respect to the
brackets the algebra $so(2,1)$
(\ref{jj}).
This problem was considered
in the paper \cite{sl2r} in connection with the quantization
of the models of relativistic particles with curvature and torsion
\cite{pl5,kuz},
and here it is reasonable to sketch the results necessary for the
present consideration.

As we have pointed out in the beginning of the section,
there are two cases being
essentially different from the point of view of the
topology of the subspace described by the variables $J_\mu$.
This topology is defined by the value of the constant $C$.
First, let us consider the case when $C=-\alpha^{2}$, $\alpha>0$.
Here we have two disconnected sheets of the hyperboloid
(\ref{2hyp}). In the quantum case the operators $\hat{J}_\mu$
can be realized in the form
\begin{equation}
\hat{J}_0=z\frac{\partial}{\partial z}+\alpha,\quad
\hat{J}_1=-\frac{1+z^2}{2}\frac{\partial}{\partial z}-\alpha z,\quad
\hat{J}_2=-i\frac{1-z^2}{2}\frac{\partial}{\partial z}+i\alpha z.
\label{d+}
\end{equation}
These operators act on the space of functions
$\psi(z)$, holomorphic in the unit disc $\vert z\vert<1$
on the complex plane.
They satisfy the commutation relations
\begin{equation}
[\hat{J}{}^{\mu},\hat{J}{}^{\nu}]=
-i\epsilon^{\mu\nu\lambda}\hat{J}_{\lambda}
\label{hjj}
\end{equation}
corresponding to the classical relations (\ref{jj}). Moreover,
the Casimir operator of the $so(2,1)$ algebra (\ref{hjj})
takes here the value
\begin{equation}
\hat{J}{}^2=-\alpha(\alpha-1)
\label{casd+}
\end{equation}
substituting the classical value $C=-\alpha^2$.
The scalar product in the internal subspace,
\begin{equation}
(\psi_1,\psi_2)=\frac{2\alpha-1}{2}\int\int_{\vert z\vert<1}
\overline{\psi_1(z)}\psi_2(z)(1-\vert z\vert^2)^{2\alpha-2}d^2z,
\label{scd+}
\end{equation}
where bar means a complex conjugation,
is defined so that the operators (\ref{d+}) are hermitian.
The infinite set of functions
\begin{equation}
\psi^n(z)=\sqrt{\Gamma(2\alpha+n)/\left(\Gamma(n+1)\Gamma(2\alpha)\right)}
\cdot z^n,\quad
n=0,1,2,\ldots,
\label{eizn}
\end{equation}
represents a complete set of orthonormal functions on the space of the
holomorphic functions, which are the eigenfunctions of the operator
$\hat{J}_0$:
\begin{equation}
\hat{J}_0\psi^n=j^n_0\psi^n,\quad
j^n_0=\alpha+n.
\label{j0+}
\end{equation}
So,  relations (\ref{casd+}) and (\ref{j0+}) mean that
the operators $\hat{J}_\mu$ (\ref{d+}) correspond to the
realization of the infinite-dimensional unitary
discrete-type series of
representations $D^{+}_{\alpha}$ of the group
${\rm \overline{SL(2,R)}}$ on the space of holomorphic functions
in the unit disc $\vert z\vert <1$ \cite{barg}.
Such a realization  of the operators
$\hat{J}_\mu$ corresponds, in turn,
to the classical variables $J_\mu$ setting the
upper sheet ($\varepsilon=+$)
of the classical two-sheet hyperboloid (\ref{2hyp}).
The realization corresponding to the lower sheet with $\varepsilon=-$
can be obtained from the realization (\ref{d+})
through the obvious  substitution:
\begin{equation}
\hat{J}_0\rightarrow -\hat{J}_0,\quad
\hat{J}_1\rightarrow -\hat{J}_1,\quad
\hat{J}_2\rightarrow \hat{J}_2.
\label{d-}
\end{equation}
In this case operators $\hat{J}_\mu$
satisfy relations (\ref{hjj}) and (\ref{casd+}), whereas
functions (\ref{eizn}) are the eigenfunctions of the operator $\hat{J}_0$
with the eigenvalues $j^n_0=-(\alpha+n)$, i.e. for $\varepsilon=-$
in eq. (\ref{2hyp}) we arrive at the unitary irreducible representations
$D^-_\alpha$ of the group ${\rm \overline{SL(2,R)}}$.
In the general case (\ref{2hyp})
we have a direct sum of the representations (\ref{d+})
and (\ref{d-}): $D^+_\alpha\oplus D^-_\alpha$.

In the case  $C=\beta^{2}\geq 0$, $\beta\geq 0$,
the quantization of the subsystem described by the vector  $J_\mu$
results in the irreducible
unitary infinite-dimensional representations of the principal
continuous series $C^{\vartheta}_{\sigma}$,
characterized by the value of the Casimir
operator
\begin{equation}
\hat{J}{}^{2}=\sigma=\beta^{2}+1/4
\label{cts}
\end{equation}
and by the eigenvalues
\begin{equation}
j^n_{0}=\vartheta+n, \quad \vartheta\in [0,1),\quad
n=0,\pm1,\pm2,\ldots,
\label{eigth}
\end{equation}
of the operator $\hat{J}_0$.
In this case the operators $\hat{J}_\mu$
are realized in the form of the linear differential operators
\begin{equation}
\hat{J}_0=-i\frac{\partial}{\partial \varphi}+\vartheta,\quad
\hat{J}_{\pm}=J_1\pm iJ_2=e^{\pm i\varphi}\left(
-i\frac{\partial}{\partial \varphi}+\vartheta\pm
i\left(\beta -\frac{i}{2}\right)\right)
\label{dcth}
\end{equation}
acting on the space of
functions  $\psi(\varphi)$ being $2\pi$-periodic in the angle variable
$\varphi$.
These operators are hermitian ones with respect to the natural
internal scalar product
\begin{equation}
(\psi_1,\psi_2)=\frac{1}{2\pi}
\int^{2\pi}_{0}\overline{\psi_1}(\varphi)\psi_2(\varphi)d\varphi,
\label{sph}
\end{equation}
and here the set of the functions
\begin{equation}
\psi^n(\varphi)=e^{in\varphi}
\label{sphi}
\end{equation}
forms the complete orthonormal
set of eigenfunctions of the operator $\hat{J}_0$
corresponding to the eigenvalues
(\ref{eigth})\footnote{With the help of the formal substitution
$\beta\rightarrow i\tilde{\beta}$,
representation (\ref{dcth})
can be transformed into the unitary irreducible
representation of the supplementary
continuous series for the case when
$0<\tilde{\beta}<1/2$. In this case
the scalar product can be changed in such a way that the modified operators
$\hat{J}_\mu$ will be hermitian ones
(see refs. \cite{sl2r,barg,muk}).}:
\[
\hat{J}_0\psi^n=j^n_0\psi^n.
\]

Here the following remark is in order. In the cases of the half-bounded
representations $D^\pm_\alpha$ and unbounded
representations of the principal continuous series $C^\vartheta_\sigma$
the operators $\hat{J}_\mu$ act on the
spaces of single-valued functions being the
holomorphic or $2\pi$-periodic functions,
respectively. The differential operator $\hat{J}_0$ contains in both cases
a constant number addition being equal to $\pm\alpha$ or $\vartheta$.
One can remove this constant addition via the appropriate transformation of
the operators $\hat{J}_\mu$, not violating either the commutation relations
or the value of the Casimir operator and eigenvalues of $\hat{J}_0$.
In the case of representation (\ref{d+}) the corresponding
transformation has the form
\begin{equation}
\hat{\tilde J}_\mu=z^\alpha\hat{J}_\mu z^{-\alpha},
\label{trd+}
\end{equation}
whereas in the case (\ref{dcth}) it is represented as
\begin{equation}
\hat{\tilde J}_\mu=e^{i\vartheta\varphi}\hat{J}_\mu e^{-i\vartheta\varphi}.
\label{trth}
\end{equation}
In the first case the eigenfunctions (\ref{eizn}) are substituted by
the functions $\tilde{\psi}{}^n(z)=z^\alpha\psi^n(z)$,
whereas in the second case the eigenfunctions
(\ref{sphi}) are substituted by
$\tilde{\psi}{}^n(\varphi)=
e^{i\vartheta\varphi}\psi^n(\varphi)$. In both cases, generally,
these transformed functions are multi-valued ones,
and we note that such multi-valuedness of the
functions taking place,
as will be shown just below, for the
one-particle relativistic massive states with fractional spin
here corresponds to the multi-valuedness of the wave functions describing
two-particle systems of nonrelativistic anyons
in one and two space dimensions \cite{multi}.
Moreover, it is also interesting to note that the
half-bounded representations $D^{\pm}_{\alpha}$
take place as representations of the algebra of observable
operators for the system of two nonrelativistic anyons
in one-dimensional space \cite{multi2}.

Let us consider now a field
$\Psi(x,\cdot)=\sum_{n}\Psi^{n}(x)\psi^n(\cdot)$,
being transformed
according to one of the described unitary infinite-dimensional
representations, which satisfies the equations:
\begin{equation}
(\hat{p}{}^{2}+m^{2})\Psi=0,\quad
(\hat{p}\hat{J}-sm)\Psi=0.
\label{maj}
\end{equation}
Here we suppose that dot means $z$ or $\varphi$, and $n=0,1,2,\ldots$,
in the first case, whereas in the second case $n=0,\pm1,\pm2,\ldots$.
These equations are
the quantum analogs of the classical constraints
(\ref{p2m2}) and (\ref{pjsm})
fixing the values of the corresponding Casimir operators
$-\hat{p}{}^{2}$ and $\hat{S}$ of the Poincar\'e group $\overline{ISO(2,1)}$.
The equations have nontrivial solutions under the appropriate
choice of the parameter $s$ coordinated with
the choice of the corresponding representation of
${\rm \overline{SL(2,R)}}$ characterized
by the parameters $\alpha$ or $\vartheta$. The simplest way to get convinced
in the validity of this statement consists in passing over to
the momentum representation and choosing the rest
frame system with ${\bm p}={\bf 0}$.
In particular,
in the case $C=-\alpha^{2}$, $s=\alpha$, we have the
discrete type representations $D^{\pm}_{\alpha}$, and the states
$\Psi\propto \psi^{0}(z)$ will describe the
physical states with positive ($p^{0}=m$) and negative ($p^{0}=-m$)
energies in the rest frame system for the cases of representations
$D^{+}_{\alpha}$  and $D^{-}_{\alpha}$,
respectively.
So, we conclude that the quantized minimal extended model corresponds
to the approach using infinite-dimensional unitary representations of the
$\overline{SL(2,R)}$ group \cite{pl5,jn1}.

\section{Anyon as a coupled two-particle system}
This section is devoted
to the consideration of the model proposed in ref. \cite{pl3}.
Here we shall get another form for its lagrangian, that will allow
us to simplify considerably the analysis of the system and
reveal a nontrivial relationship of the model to the minimal
extended model from the preceding section.  We shall also
comment on the relationship of the approach of Balachandran et
al. \cite{bal} to the present one and we shall point out
an interesting possible interpretation of the model
explaining the title of the section.

Thus, let us consider the system, given by
the lagrangian \cite{pl3,pl4}
\begin{equation}
L=me\dot{x}-s\epsilon_{\mu\nu\lambda}e^{\mu}n^{\nu}\dot{n}{}^{\lambda}
-\frac{\sigma}{2}(e^{2}+1)-\frac{\rho}{2}(n^{2}-1)-\omega(en).
\label{len}
\end{equation}
Here $n^{\mu}$ is translationally invariant vector, which together with
corresponding conjugate momenta will play the role of the variables $z_{n}$,
whereas $\sigma$, $\rho$, $\omega$ and $e_{\mu}$ are Lagrange multipliers.
The variation of the action over these multipliers gives the
Lagrange constraints
\begin{eqnarray}
&e^{2}+1=0,\quad
n^{2}-1=0,\quad
en=0,&
\label{encon}\\
&\sigma e_{\mu}+\omega n_{\mu}+
s\epsilon_{\mu\nu\lambda}n^{\nu}\dot{n}{}^{\lambda}-m\dot{x}_{\mu}=0.&
\label{emu}
\end{eqnarray}
Using constraints (\ref{emu}) alongside with the
first and third ones from the set (\ref{encon}),
we can express the Lagrange
multipliers $\sigma$, $\omega$ and $e_{\mu}$ via $\dot{x}_{\mu}$,
$n_{\mu}$ and $\dot{n}_{\mu}$. Then,
putting their expressions into
lagrangian (\ref{len}), we get for it the following form,
equivalent from the physical point of view
to the initial one:
\begin{equation}
L=-m\sqrt{-(\dot{x}_{\mu}-n_\mu (\dot{x}n)-sm^{-1}\epsilon_{\mu\nu\lambda}
n^{\nu}\dot{n}{}^{\lambda})^{2}}-\frac{\rho}{2}(n^{2}-1).
\label{lxn}
\end{equation}
This lagrangian leads to the mass shell constraint
(\ref{p2m2}), the spin constraint (\ref{pjsm}) (see below) and
the constraint $\Pi\approx 0$
as the first class constraints. Moreover, there is a set
of second class constraints in the system:
\begin{eqnarray}
&pn\approx 0,\quad p\pi\approx 0,&
\label{sec1}\\
&\pi n\approx 0,\quad n^{2}-1\approx 0.&
\label{sec2}
\end{eqnarray}
Here by $\Pi$ and
$\pi_{\mu}$ we denote the momenta conjugate to $\rho$ and $n^{\mu}$,
\begin{equation}
\{\rho,\Pi\}=1,\quad
\{n_{\mu},\pi_{\nu}\}=\eta_{\mu\nu}.
\label{qpi}
\end{equation}
The constraint $\Pi\approx 0$ means that
$\rho$ and $\Pi$ are pure gauge variables. They can be
eliminated, e.g., by introducing the gauge $\rho-\rho_{0}\approx
0$ to this constraint, where $\rho_0$ is some constant.  At the
same time, the second class constraints (\ref{sec1}) and
(\ref{sec2}) have the following sense: the first pair serves
for removing the nonphysical degree of freedom described by the
timelike conjugate variables $n^{(0)}= ne^{(0)}$ and
$\pi^{(0)}=\pi e^{(0)}$, $\{n^{(0)},\pi^{(0)}\}=-1$, whereas
constraints (\ref{sec2}) remove (or freeze according to our
terminology from sections 2 and 4) the radial oscillator degree
of freedom described by the conjugate variables
$n_r=\sqrt{n^{\bot 2}}$ and $\pi_r=\pi^{\bot}n$,
$\{n_r,\pi_r\}=1$, being
independent in the sense of brackets from $n^{(0)}$ and $\pi^{(0)}$.
These four variables play, obviously, the
role of the variables $\tilde{z}_a$ discussed in section 4.

In the present model the total angular momentum has
the form given by eq. (\ref{calj}) with
spin addition equal to
\begin{equation}
J_{\mu}=-\epsilon_{\mu\nu\lambda}n^{\nu}\pi^{\lambda}.
\label{jnp}
\end{equation}
It is this vector which defines the spin constraint (\ref{pjsm}) here.
Vector (\ref{jnp}) satisfies the $so(2,1)$ algebra (\ref{jj}) with
respect to the initial canonical Poisson brackets of $x_{\mu}$, $p_{\mu}$,
$n_{\mu}$ and $\pi_{\mu}$, but it has nonzero Poisson brackets with the
constraints (\ref{sec1}).  This means that after necessary taking into
account second class constraints, the ``spin vector" (\ref{jnp}) will satisfy
another (trivial) algebra with respect to the corresponding Dirac brackets,
and, moreover, as we shall see,
it will turn out to be parallel to the
energy-momentum vector $p_\mu$  on the surface of the second class
constraints (\ref{sec1}).

Thus, let us take into
account the second class constraints (\ref{sec1}), (\ref{sec2}),
that will allow us to exclude from the consideration
nonphysical degrees of freedom described by the variables
$n^{(0)}$ and $n_r$ and by their conjugate momenta.
The simplest way to do this consists in reducing the system to
the surface $\Gamma$ specified by
these constraints.  On $\Gamma$, the
variables $n_{\mu}$ and $\pi_{\mu}$ can be parametrized in the form:
\begin{equation}
n_{\mu}=e^{(i)}_{\mu}n^{(i)},\qquad \pi_{\mu}=J^{(0)}e^{(i)}_{\mu}
\epsilon^{ij}n^{(j)},
\label{gama}
\end{equation}
where $e^{(i)}_{\mu}$, $i=1,2$, are two components of the triad
(\ref{triad}), $n^{(i)}=(\sin\varphi,\cos\varphi)$, $0\leq\varphi<2\pi$, and
the quantity $J^{(0)}$,
$-\infty<J^{(0)}<\infty$, is understood as an independent variable
on the reduced phase space of the system.
 Parametrization (\ref{gama}) leads to the equality
$J_{\mu}=-J^{(0)}e^{(0)}_{\mu}$.  Therefore, we see that
on the surface of the second class constraints, $J_\mu$ is indeed
parallel to the vector $p_\mu$ and the quantity $J^{(0)}$ has, according to
the definition (\ref{spin}), a sense of the spin variable.  Then, reducing
the symplectic 2-form $\Omega=dp_{\mu}\wedge dx^{\mu}+ d\pi_{\mu}\wedge
dn^{\mu}$ to $\Gamma$, we get the symplectic 2-form
\[
\omega=dp_{\mu}\wedge dy^{\mu}+dJ^{(0)}\wedge d\varphi
\]
on the reduced phase space described by the variables
$J^{(0)}$, $\varphi$, $p_{\mu}$ and
\begin{equation}
y_{\mu}=x_{\mu}-J^{(0)}e^{(1)\nu}\partial_{\mu} e^{(2)}_{\nu}.
\label{zmu}
\end{equation}
{}From this 2-form we find that the variables of the
reduced phase space have the following nonzero Dirac brackets:
\begin{equation}
\{y_{\mu},p_{\nu}\}^*=\eta_{\mu\nu},
\label{zp}
\end{equation}
\begin{equation}
\{\varphi,J^{(0)}\}^*=1.
\label{fij}
\end{equation}
With the help of eq. (\ref{zmu})
one can find nontrivial Dirac brackets for the initial coordinates~$x_{\mu}$:
$\{x_\mu,p_\nu\}^*=\eta_{\mu\nu},$
\begin{equation}
\{x_{\mu},x_{\nu}\}^{*}=-J^{(0)}\epsilon_{\mu\nu\lambda}
\frac{p^{\lambda}}{(-p^{2})^{3/2}},
\label{xxd}
\end{equation}
\begin{equation}
\{x_{\mu},\varphi\}^{*}=-e^{(1)\sigma}\partial_{\mu}e^{(2)}_{\sigma}.
\label{xfd}
\end{equation}
The Dirac brackets for the initial ``internal" variables $n_\mu$ and
$\pi_\mu$ between themselves and with the coordinates $x_\mu$
can be found with the help of the parametrization (\ref{gama}) and
corresponding Dirac brackets of the reduced phase space variables.

Therefore, we can describe the reduced system in terms of the
variables $J^{(0)}$, $\varphi$, $p_\mu$ and $x_\mu$
(or $y_\mu$ instead of $x_\mu$). The total
angular momentum has here the form given by eq. (\ref{caljtil})
(with $x_\mu$ instead of $\tilde{x}_\mu$),
whereas the spin constraint (\ref{pjsm}) is presented now in the form
(\ref{j0s}).  Comparing the form of the brackets
(\ref{fij}), (\ref{xxd}) and (\ref{xfd}) with the brackets
(\ref{vj1}), (\ref{txtx})
and (\ref{txf}), respectively, one can conclude that we have arrived exactly
at the system described in section 4 within a framework of the
general analysis of the extended canonical formulation in the case
when $2N=2$.
That corresponding system comprises
the variables $\tilde{x}_{\mu}$, $p_{\mu}$, $J^{(0)}$ and $\varphi$,
whereas the variables $\tilde{z}_a$ have been excluded already by us.

In order to quantize the system, we have to realize the
operators corresponding to the coordinates $x_\mu$,
nonlocalizable in a sense of the Dirac brackets (\ref{xxd}).
It can be done as in the case of the minimal approach with the
help of the localizable but noncovariant coordinates $y_\mu$,
whose noncovariance is conditioned by the noncovariant second
term in eq. (\ref{zmu}).
We can choose a representation with the operators $\hat{y}_\mu$
being diagonal, and  work on the space of the $2\pi$-periodic
wave functions
\begin{equation}
\Psi(y,\varphi)=\sum e^{i\tilde{n}\varphi}\psi^n(y).
\label{periodic}
\end{equation}
Here we suppose that $\tilde{n}=n$ and
summation is realized over $n\in Z$.
In accordance
with the classical equalities (\ref{zp}), (\ref{fij}), we realize
other operators as differential operators:
$\hat{p}_\mu=-i\partial/\partial y^{\mu}$ and
$\hat{J}{}^{(0)}=-i\partial/\partial\varphi +\vartheta$, where
$\vartheta$ is an arbitrary constant,
whose value can be restricted to the region
$\vartheta\in [0,2\pi)$ without loss of generality.
Under a choice of the internal scalar
product in the form (\ref{sph}), the operator $\hat{J}{}^{(0)}$
is a hermitian operator for any value of the constant
$\vartheta$.  As in the case of the minimal approach, in this
representation the operators corresponding to the initial
coordinates $x_\mu$, realized with the help of the quantum
analog of eq. (\ref{zmu}), turn out to be nonlocal operators.
The quantum analogs of the mass shell and spin constraints turn
into equations which single out the physical subspace of the
system, and we find that the spin equation,
$(\hat{J}{}^{(0)}-s)\Psi=0$, has nontrivial solutions when the
parameter $\vartheta$ is chosen to be equal to the fractional
part of the spin parameter $s$.  As a result, we find that the
physical states are described by the wave functions of the form
\[
\Psi_{phys}(y,\varphi)
\propto\int d^3 p e^{ipy}\delta(p^2+m^2)e^{i[s]\varphi},
\]
where $[s]$ is the integer part of the spin parameter $s$.
On the other hand, as in the case of the quantum scheme for the
model considered in the previous section,  we can realize
the operator $\hat{J}{}^{(0)}$ in the standard form, without constant
shift $\vartheta$,
$\hat{J}{}^{(0)}=-i\partial/\partial\varphi$.
In this case the spin equation will have nontrivial solutions if
we choose the space of functions as that which is defined by the
decomposition of the form (\ref{periodic}), but with $\tilde{n}$
to be equal to $n+(s-[s])$. As a result, the physical wave functions
will contain a corresponding
factor of the form $e^{is\varphi}$ being multi-valued in general case.
Thus, the described scheme of quantization
reveals a similarity with that
for the minimal approach and with the quantum scheme
for the minimal extended model
from the previous section for the case
$C\geq 0$.
The latter similarity forces us to investigate in more detail the
relationship between the two extended models.
Below we shall show that the reduced phase space description of
the present model can be put in one-to-one correspondence with
the canonical description of the previous model specified by the value of
the constant $C\geq 0$.

To reveal such a relationship, let us note
that here the  internal phase space described
by the variables  $\varphi$ and $J^{(0)}$ having the
Dirac brackets (\ref{fij}) is a cotangent bundle $T^*S^1$ of
the one-dimensional sphere $S^1$.
Topologically, this space is a cylinder, which, in turn,
is equivalent to the one-sheet two-dimensional hyperboloid.
Thus, let us construct the following vector on the reduced phase space
of the system:
\begin{equation}
J^{\star}_\mu=-\epsilon_{\mu\nu\lambda}n^{\nu}\pi^{\lambda}+
\sqrt{C}n_\mu+\pi_\mu.
\label{jstar}
\end{equation}
It has the fixed square, $J^{\star}_\mu J^{\star\mu}=C$,
characterized by the constant $C$, which is supposed to be a nonnegative one,
$C\geq 0$. In correspondence with eq. (\ref{gama}),
$J^{\star}_\mu e^{(0)\mu}=J^{(0)}$
and $J^{\star}_\mu e^{(i)\mu}=(\sqrt{C}\delta^{ij}+J^{(0)}
\epsilon^{ij})n^{(j)}$, and we conclude that eq. (\ref{jstar})
gives a map of the cylinder
to the one-sheet two-dimensional hyperboloid (\ref{const})
with $C\geq 0$.

Now, let us note that the spin constraint (\ref{pjsm})
has here exactly the same form in terms of the quantities
$J^{\star}_\mu$.  Moreover,
with the help of the parametrization of the reduced phase
space (\ref{gama}) and corresponding Dirac brackets,
one can check that $J^{\star}_\mu$ satisfies the $so(2,1)$ algebra
of the form (\ref{jj}), whereas here
the Dirac brackets of $x_\mu$ with $J^{\star}_\nu$ have the form
\[
\{x_\mu,J^{\star}_\nu\}^*=\frac{1}{p^2}\left(\eta_{\mu\nu}(pJ^{\star})-
J^{\star}_\mu p_\nu\right)
\]
coinciding
with the form of the corresponding
Poisson brackets of the gauge-invariant coordinates
$\tilde{x}_\mu$, given by eq. (\ref{tildx}), with $J_\mu$ (see section 4).
Then, due to the coincidence of the form of the Dirac brackets
(\ref{xxd}) for
the coordinates $x_\mu$ with the form of the Poisson
brackets (\ref{txtx}) for the coordinates $\tilde{x}_\mu$,
we arrive at the conclusion that the new  coordinates
\begin{equation}
x^{\star}_\mu=x_\mu-\frac{1}{p^2}\epsilon_{\mu\nu\lambda}p^{\nu}
J^{\star\lambda}
\label{xstar}
\end{equation}
have the trivial Dirac brackets with $J^{\star}_\nu$,
$\{x^{\star}_\mu,J^{\star}_\nu\}^*=0$, and
\begin{equation}
\{x^{\star}_\mu,x^{\star}_\nu\}^*=0,
\label{xxstar}
\end{equation}
whereas the form of the total angular momentum vector of the system
under consideration
is presented in terms of $x_\mu^{\star}$, $p_\mu$ and
$J^{\star}_\mu$ in the same form (\ref{calj}) as the total
angular momentum vector of the minimal extended model from section 5.
Therefore, both
$x^{\star}_\mu$ and $J^{\star}_\mu$  are Lorentz vectors,
and we conclude that we indeed have established a one-to-one
correspondence of the reduced phase space description of the
present model with the minimal
extended model for the case when $C\geq 0$.
At the same time, it is necessary to
note that the reduced phase space of the system under consideration
is topologically different from
the phase space of the minimal extended model in the case when
there constant $C$ is chosen to be negative.

Let us comment on a possible interpretation of the model that has been
considered.
Such an interpretation will explain the title of the present section.
For the purpose we note that
the second term in the lagrangian (\ref{lxn}) prescribes the spacelike
vector $n_\mu$, $n^2>0$, to have  a fixed (Minkowsky) length, $n^2=1$.
One could omit this term from the lagrangian if one
present $n_\mu$ in the first term as $n_\mu=q_\mu/\sqrt{q^2}$,
where $q^\mu$ is supposed to be an arbitrary spacelike vector.
The only difference which we have in this case is the
absence of the second constraint from the pair of constraints (\ref{sec2}),
whereas other constraints from the set of constraints (\ref{sec1})
and (\ref{sec2}) are substituted here by $pq\approx 0$, $p\pi\approx 0$,
and $\pi q\approx 0$,
where now $\pi_\mu$ is the canonical momenta conjugate to $q^{\mu}$.
In this case the last constraint means
simply that the length of the spacelike vector
$q_\mu$ is an unobservable quantity
(and it can be fixed, e.g., by introducing the constraint
$q^2-1\approx 0$ as a gauge condition to the last constraint).
This guarantees that the physical content of the model is
not changed.
Since $q_\mu$ is a translationally invariant vector,
we can introduce the vectors $x_{a}^{\mu}$, $a=1,2$,
via the definition
\[
x_\mu=\frac{1}{2}(x_{1\mu}+x_{2\mu}),\quad
q_\mu=x_{1\mu}-x_{2\mu}.
\]
Then, in the rest frame of the system defined by the equality
${\bm p}={\bf 0}$,  the constraint $pq\approx 0$ is reduced
to $x_1^{0}=x_2^0$, and, therefore, here the condition that the vector
$q_\mu$ is a spacelike one means that ${\bm x}_1\neq {\bm x}_2$.
Thus, we can interpret the system as the system of two coupled relativistic
particles described by the space-time coordinates $x^\mu_1$ and $x^\mu_2$,
having a nontrivial
configuration space defined by the condition $(x_1-x_2)^2>0$, which
prescribes the particles do not have coinciding space coordinates.
In such an interpretation the system is related
to the nonrelativistic system of two anyons discussed by
Leinaas and Myrheim in their
first original paper on the subject of fractional statistics \cite{leimyr}.
Then the form of lagrangian of the model under consideration
guarantees that such a system of coupled particles has only
internal spin degree of freedom and the values of its
corresponding relativistic mass and spin are fixed.

To conclude this section,
we point out how the model considered here is related to the
approach of Balachandran et al. \cite{bal} translated to
the (2+1)-dimensional space-time.  Proceeding from the original
form of the lagrangian (\ref{len}), one can supply the
system of the two vectors $e_{\mu}$ and $n_{\mu}$ with a
third one, $l_\mu=\epsilon_{\mu\nu\lambda} e^\mu n^\lambda$.
Then, taking into account lagrangian constraints
(\ref{encon}), we see that these vectors
form the complete orthonormal set of vectors, $-e_\mu e_\nu+n_\mu
n_\nu+l_\mu l_\nu=\eta_{\mu\nu}$.
Now, from these vectors we can compose the
matrix $\Lambda^{\mu\nu}$, $\Lambda^{\mu 0}=e^{\mu}$,
$\Lambda^{\mu 1}=n^{\mu}$, $\Lambda^{\mu 2}=l^{\mu}$,
which, obviously, is the matrix of a Lorentz transformation:
$\Lambda^{\mu\lambda}\Lambda_{\nu\lambda}=\eta^\mu_\nu$.
Such a matrix as a dynamical object is the main ingredient
of the above-mentioned approach,
where it is introduced for taking into account the spin
degrees of freedom.  Thus, the approach of the present model is
related to the approach \cite{bal}.

\section{Summary and conclusions}

We have shown that, within the framework of the minimal
canonical approach, (2+1)-dimensional relativistic massive
spinning particle can be described by the coordinates
$x_\mu$ generally having nontrivial brackets given by eqs. (\ref{xxr})
and (\ref{rmua}). The form of the total angular momentum (\ref{caljm})
of the particle with spin $s$
is correlated with the form of the brackets.  Due to the presence
of the term with specific dependence on the ``gauge field" $A_\mu(p)$,
the spin addition $J_\mu=-se^{(0)}_\mu
-\epsilon_{\mu\nu\lambda}A^\nu p^\lambda$
in the total angular momentum vector (\ref{caljm}) generally is not parallel
to the energy-momentum vector of the particle $p_\mu$.
Besides, in general case the coordinates $x_\mu$ have transformation
properties different from those for a Lorentz vector.
But there are two special cases here. One case is given by the trivial
(up to the ``gauge transformation" (\ref{gt}))
``gauge field" $A_\mu=0$.
{\it Only in this case the corresponding
coordinates $x^{c}_\mu$ form a Lorentz vector
and the spin addition $J^{c}_\mu=-se^{(0)}_\mu$
turns out to be parallel to the vector $p_\mu$.} But the price
which we have to pay for the covariant properties of
$x^{c}_\mu$ is the specific nontrivial form of their brackets
(\ref{xxc}) meaning that at the quantum level there is no
representation where the corresponding operators
$\hat{x}{}^{c}_\mu$ would be diagonal.  It is necessary to
stress that here either the first, orbital-like term in eq.
(\ref{caljc}), or the ``spin vector" $J^{c}_\mu$ does not form
the Lorentz algebra $so(2,1)$ with respect to the brackets,
but only the total angular momentum vector ${\cal J}_\mu$ does.
Moreover, let us stress that
within the framework of the minimal formulation we always have
the equality  $\{J_\mu,J_\nu\}=0$.
Another special case is characterized by the localizable coordinates
$x^{l}_\mu$ having trivial brackets (\ref{local}).
This case is given by the SO(2,1)
``monopole gauge potential" defined by eq. (\ref{monopol}), whose
solution, in turn, is given by eqs. (\ref{del}) and (\ref{xif}).
Here the spin addition $J^{l}_\mu$ is not parallel
to the energy-momentum vector $p_\mu$, and
localizable coordinates $x^{l}_\mu$ have noncovariant
transformation properties with respect to the Lorentz boosts.
We have demonstrated that {\it the latter, noncovariant,
special case turns out to be important
for the construction of the quantum theory corresponding
to the former case given by the covariant but nonlocalizable
coordinates.}

$\ $

{\it The properties of covariance and localizability of the
coordinates can be simultaneously incorporated into the theory
within the framework of the extended
canonical formulation.} Such a compatibility is achieved via
introducing auxiliary internal phase space variables and
by fixing the value of the particle spin in the weak sense, with the help
of the spin constraint. Here the components of the
spin addition $J_\mu$ form the $so(2,1)$ algebra
with respect to the Poisson brackets,
$\{J_\mu,J_\nu\}=-\epsilon_{\mu\nu\lambda}J^{\lambda}$,
but, generally, this vector is not
parallel to  the energy-momentum vector $p_\mu$ in the weak sense,
on the surface of the spin constraint.
One  can reformulate the extended approach
in terms of variables which are gauge-invariant
with respect to the gauge transformations generated by the spin constraint,
presented in the dimensionless form (\ref{j0s}).
Such a reformulated extended canonical approach
turns out to be equivalent to the minimal approach
given in terms of the covariant nonlocalizable coordinates.
On the other hand, the general case of the minimal formulation
can be obtained from the extended formulation via reducing
the latter one to the surface of the spin constraint.
We have shown that there is only one special case, when
the gauge-invariant extension $\tilde{x}_\mu$ of the
coordinates of the particle, having the brackets
corresponding to the brackets of the covariant coordinates
$x^{c}_\mu$, coincides with the initial coordinates $x_\mu$.
This case is characterized by the spin vector $J_\mu$ being
parallel to the energy-momentum vector $p_\mu$
on the surface of the spin constraint.
Here the gauge orbits generated by the spin constraint (\ref{j0s}),
which generally are one-dimensional spheres,
shrink into just one point.

Such a special case is contained as a particular one in the
concrete minimal extended model has been considered in
section 5.  We have shown that this case turns out also to be
a special one from the point of view  of the lagrangian
approach.  Only in this case there is the Lagrange constraint in
the system prescribing the velocity of the particle to be
parallel to the spin vector $J_\mu$, and,
therefore, only in this case the energy-momentum vector of the
particle, $p_\mu$, is parallel to its velocity $\dot{x}_\mu$.
In the general case the coordinates of the particle are subject
to the more complicated motion being a superposition of the
circular  and rectilinear motion along the momentum vector ${\bm p}$.
Such a motion represents by itself the classical analog of the
quantum Zitterbewegung \cite{zit}, and, so, {\it generally the
second derivative of the coordinates of the particle
(acceleration in the laboratory temporal gauge) is not equal to
zero for a free relativistic spinning particle in 2+1
dimensions.} At the same time, the above-mentioned
{\it gauge-invariant extension $\tilde{x}_\mu$ of the
coordinates $x_\mu$ does not reveal any circular motion and its
evolution law has the form (\ref{foldi}) exactly coinciding
with that for the evolution of the free relativistic spinless
particle.}  Thus, the coordinates $\tilde{x}_\mu$ turn out to
be analogous to the Foldy-Wouthuysen coordinates for the
quantum massive Dirac particle \cite{foldy}.

Let us stress once more that in the special case of the model
with the spin parameter correlated with the value of the
central element of the $so(2,1)$ algebra (\ref{jj}),
$s^2=-J^2$, the spin addition $J_\mu$ is parallel to $p_\mu$ on the
surface of the spin constraint, the velocity of the particle
is parallel to energy-momentum vector $p_\mu$, and so,
as we have seen, the
coordinates of the particle evolve in such a way that
$\ddot{x}_\mu=0$.  At the same time, {\it we have here the
Poincar\'e group as the exact symmetry group of the system.}
Therefore, the general statement of the paper \cite{dutra},
declaring the
incompatibility of the free nature of the anyon (characterized there
by the relation $\ddot{x}_\mu=0$) and translation invariance of the
theory, turns out to be incorrect one.

We have pointed out that after reduction of the system to the
surface of the spin constraint in the above-mentioned special
case the spin addition $J_\mu$ turns out to be parallel
to $p_\mu$ in the strong sense,
but its components, in correspondence with the
general results obtained for the minimal approach,
form the trivial algebra with respect to the Dirac brackets.
The quantization of the model in this special case leads to the
representations of the discrete type series $D^{\pm}_\alpha$ of
the group $\overline{\rm SL(2,R)}$, characterized by
the parameter $\alpha$ being correlated with the spin of the
particle, $s^2=\alpha^2$.
Let us note here that different variants of linear differential
equations for fractional spin fields,
have been constructed up to now
within a group-theoretical approach to anyons, are known
only for the case of using such
representations of the discrete type series \cite{cor,pl3,jn1,pl6}.
Moreover, it is necessary to note that
there is a vector set of linear differential equations,
proposed in ref. \cite{cor}, which itself fixes the choice of only
these representations for the description of fractional spin
fields.  The quantum analogs of the mass shell (\ref{p2m2}) and
spin constraints (\ref{pjsm}) appear there as a consequence of the
corresponding basic linear differential equations.

We have shown that in the case when the parameter $C$, defining the
topology of the internal phase space of the
minimal extended model, is nonnegative, the model turns out, in fact,
to be equivalent to the model of refs. \cite{pl3,pl4}.
In turn, the latter model can be interpreted as the system of
two coupled relativistic particles with nontrivial topology of
configuration space, which prohibits the particles to have coinciding
space coordinates.  The form of the lagrangian of the model guarantees
removing all the degrees of freedom corresponding
to the relative motion of the particles being
different from the spin degree of freedom, and, moreover,
it prescribes the system to have fixed relativistic total mass
and spin.
We have also demonstrated that the model from section 6 is related
to the approach of ref. \cite{bal}, transferred to the case of 2+1
dimensions.

$\ $

Finally, let us list the main results removing
the misleading  notions on the general properties of
the (2+1)-dimensional anyons which have been cited in section 1.

$\ $

{\it A free relativistic particle with fractional (arbitrary)
spin can be described in 2+1 dimensions in a
Poincar\'e-invariant way.  Under such a description, the
classical analog of the relativistic quantum Zitterbewegung
generally takes place, and, so, the condition $\ddot{x}_\mu=0$
is not a necessary condition for a particle to be free.

The spin addition $J_\mu$ of the total angular momentum vector
cannot simultaneously satisfy the properties of parallelness
to the energy-momentum vector $p_\mu$ and $so(2,1)$ algebra
$\{J_\mu,J_\nu\}=-\epsilon_{\mu\nu\lambda}J^{\lambda}$.
$J_\mu$ can be (but is not necessarily) parallel to $p_\mu$ in the
strong sense within the framework of the minimal formulation,
where its components satisfy the trivial algebra
$\{J_\mu,J_\nu\}=0$. On the other hand,
spin addition satisfies $so(2,1)$
algebra within the framework of the extended formulation,
where $J_\mu$ can be (but is not necessarily) parallel to $p_\mu$
only in the weak sense, on the surface of the spin constraint.}

$\ $

The general formulation introduced in this paper can give a new
perspective on the remaining open questions in
the study of (2+1)-dimensional anyons.  The quantum field theory
of fractional spin fields and the construction
of consistent theory of electromagnetic interaction of anyons
are two basic problems not yet solved, where the
model-independent approach presented here can be important.

$\ $

$\ $

{\bf Acknowledgements}

M.P. thanks J.M. Leinaas and J. Myrheim for useful discussions.
The work was supported by MEC-DGICYT, Spain.


\end{document}